\documentclass[a4paper,11pt]{article}
\pdfoutput=1 

\usepackage{jcappub} 

\makeatletter
\gdef\@fpheader{}
\g@addto@macro\bfseries{\boldmath}
\makeatother

\usepackage[OT1]{fontenc} 

\def\baq{\begin{eqnarray}}
\def\eaq{\end{eqnarray}}

\def\beq{\begin{equation}}
\def\eeq{\end{equation}}

\usepackage{xcolor}
\usepackage[normalem]{ulem} 
\usepackage{cancel} 

\usepackage{ifthen}
\usepackage{tikz}
\usepackage{physics,amsmath,bm}
\usepackage{xspace}
\usepackage[T1]{fontenc}

\newcommand{\GeV}{{\rm GeV}\xspace}

\newcommand{\bvec}[1]{\vectorbold{#1}\xspace}
\newcommand{\veck}{\bvec k\xspace}

\newcommand{\svev}[1]{\langle #1 \rangle}

\newcommand{\lrb}[1]{\left( #1 \right)}
\newcommand{\lrsb}[1]{\left[ #1 \right]}

\newcommand{\lrBigsb}[1]{\Big[ #1 \Big]}

\newcounter{NumArgs}

\newcommand{\eqs}[1]{\setcounter{NumArgs}{0}\foreach\i in{#1}{\stepcounter{NumArgs}}%
\ifthenelse{\equal{\theNumArgs}{1}}{eq.~(\ref{#1})}%
{\ifthenelse{\equal{\theNumArgs}{2}}%
{eqs.~\foreach\i[count=\q]in{#1}{\ifthenelse{\equal{\q}{\theNumArgs}}{and (\ref{\i})}{(\ref{\i})~}}}%
{eqs.~\foreach\i[count=\q]in{#1}{\ifthenelse{\equal{\q}{\theNumArgs}}{and (\ref{\i})}{(\ref{\i}),~}}}}}

\newcommand{\Eqs}[1]{\setcounter{NumArgs}{0}\foreach\i in{#1}{\stepcounter{NumArgs}}%
\ifthenelse{\equal{\theNumArgs}{1}}{Eq.~(\ref{#1})}%
{\ifthenelse{\equal{\theNumArgs}{2}}%
{Eqs.~\foreach\i[count=\q]in{#1}{\ifthenelse{\equal{\q}{\theNumArgs}}{and (\ref{\i})}{(\ref{\i})~}}}%
{Eqs.~\foreach\i[count=\q]in{#1}{\ifthenelse{\equal{\q}{\theNumArgs}}{and (\ref{\i})}{(\ref{\i}),~}}}}}

\newcommand{\refs}[1]{\setcounter{NumArgs}{0}\foreach\i in{#1}{\stepcounter{NumArgs}}%
\ifthenelse{\equal{\theNumArgs}{1}}{(\ref{#1})}%
{\ifthenelse{\equal{\theNumArgs}{2}}%
{\foreach\i[count=\q]in{#1}{\ifthenelse{\equal{\q}{\theNumArgs}}{and (\ref{\i})}{(\ref{\i})~}}}%
{\foreach\i[count=\q]in{#1}{\ifthenelse{\equal{\q}{\theNumArgs}}{and (\ref{\i})}{(\ref{\i}),~}}}}}

\newcommand{\Figs}[1]{\setcounter{NumArgs}{0}\foreach\i in{#1}{\stepcounter{NumArgs}}%
\ifthenelse{\equal{\theNumArgs}{1}}{Figure~\ref{#1}}%
{\ifthenelse{\equal{\theNumArgs}{2}}%
{Figures~\foreach\i[count=\q]in{#1}{\ifthenelse{\equal{\q}{\theNumArgs}}{and \ref{\i}}{\ref{\i}~}}}%
{Figures~\foreach\i[count=\q]in{#1}{\ifthenelse{\equal{\q}{\theNumArgs}}{and \ref{\i}}{\ref{\i},~}}}}}

\newcommand{\Gen}[2]{\setcounter{NumArgs}{0}\foreach\i in{#2}{\stepcounter{NumArgs}}%
	\ifthenelse{\equal{\theNumArgs}{1}}{#1.~(\ref{#2})}%
	{\ifthenelse{\equal{\theNumArgs}{2}}%
		{#1.~\foreach\i[count=\q]in{#2}{\ifthenelse{\equal{\q}{\theNumArgs}}{and (\ref{\i})}{(\ref{\i})~}}}%
		{#1.~\foreach\i[count=\q]in{#2}{\ifthenelse{\equal{\q}{\theNumArgs}}{and (\ref{\i})}{(\ref{\i}),~}}}}}

\usepackage{caption}
\usepackage{subcaption}

\usepackage{lipsum}  

\title{Stochastic Gravitational Waves from Modulated Reheating}

\author[a,b]{Michele Benaco}
\author[a,b]{Dimitrios Karamitros}
\author[a,b]{Sami Nurmi}
\author[c,b]{Kimmo Tuominen}

\affiliation[a]{Department of Physics, P.O.Box 35 (YFL), FIN-40014 University of Jyväskylä, Finland}
\affiliation[b]{Helsinki Institute of Physics, PL 64, 00014 University of Helsinki, Finland}
\affiliation[c]{Department of Physics, P.O. Box 64, FIN-00014 University of Helsinki, Finland}

\emailAdd{michele.m.benaco@jyu.fi}
\emailAdd{sami.t.nurmi@jyu.fi}
\emailAdd{kimmo.i.tuominen@helsinki.fi}

\abstract{We investigate scalar-induced stochastic gravitational waves from adiabatic curvature perturbations sourced by a spectator field via the modulated reheating mechanism. We consider a spectator scalar with Higgs-like couplings and inflaton decay via shift symmetric dimension-five operators. The spectator is  assumed to be in the Sitter vacuum and it sources blue-tilted, strongly non-Gaussian curvature perturbations which can dominate the spectrum on small scales $k \gg {\rm Mpc}^{-1}$. We find that the setup could generate a gravitational wave signal testable by surveys like BBO and DECIGO but only for large coupling values not expected in low-energy particle physics setups that can be perturbatively extrapolated up to the inflationary scale.}

\begin{document}
{\hfill{\footnotesize{HIP-2025-29/TH}}}
\maketitle

\flushbottom

\section{Introduction}
\label{sec:section1}

Energetically subdominant matter fields during inflation are commonly called spectators. Spectator scalars can for example source adiabatic or isocurvature primordial perturbations  \cite{Lyth:2001nq,Moroi:2001ct,Enqvist:2001zp,Linde:1996gt,Mollerach:1989hu,Kofman:2003nx,Dvali:2003em}, constitute non-thermal dark matter components \cite{Kuzmin:1998kk,Peebles:1999fz,Marsh:2015xka,Enqvist:2014zqa,Nurmi:2015ema,Kainulainen:2016vzv,Markkanen:2018gcw,Markkanen:2015xuw,Ema:2018ucl,Cosme:2018nly,Alonso-Alvarez:2018tus,Alonso-Alvarez:2019ixv}, affect the reheating process \cite{Figueroa:2016dsc,Dimopoulos:2018wfg,Nakama:2018gll,Opferkuch:2019zbd}, or be linked to the baryogenesis mechanism \cite{Dine:1995kz,Affleck:1984fy}. The Higgs can be a spectator field and the stability of the electroweak vacuum during inflation and reheating sets stringent constraints on its couplings \cite{Espinosa:2007qp,Lebedev:2012zw,Lebedev:2012sy,Kobakhidze:2013tn,Fairbairn:2014zia,Enqvist:2014bua,Hook:2014uia,Herranen:2014cua,Kamada:2014ufa,Espinosa:2015qea,Herranen:2015ima,Figueroa:2017slm}. Interestingly, spectator fluctuations produced during inflation or reheating could also source primordial black holes \cite{Klimai:2012sf,Young:2013oia,Kawasaki:2012wr,Carr:2016drx,Carr:2017edp,Espinosa:2017sgp,Maeso:2021xvl,Passaglia:2021jla,Stamou:2023vft,Gow:2023zzp,Chen:2024pge,Kuroda:2025coa,Garcia-Bellido:2016dkw,Pi:2021dft,Chen:2023lou} or gravitational waves \cite{Garcia-Bellido:2016dkw,Pi:2021dft,Chen:2023lou,Bartolo:2007vp,Domenech:2021and,Domenech:2023jve,Cui:2023fbg,Kumar:2024hsi,Ebadi:2023xhq, Inomata:2023drn,Garcia:2025yit,Chakraborty:2024rgl,Chakraborty:2025oyj}. 

In this work we study stochastic gravitational wave signals generated by a spectator field with a blue-tilted spectrum which sources non-Gaussian adiabatic perturbations through the modulated reheating mechanism \cite{Kofman:2003nx,Dvali:2003em}.  Such a spectator can only give a small contribution to the total curvature perturbations on large scales $k \lesssim 0.1 {\rm Mpc}^{-1}$ where perturbations are measured to be Gaussian with a red-tilted spectrum \cite{Planck:2018vyg,Planck:2019kim}. On small scales, $k \gg {\rm Mpc}^{-1}$, the spectator can however dominate the curvature perturbation and observationally testable stochastic gravitational waves could be generated if the perturbation amplitude grows large enough. Gravitational waves in a closely related setup were recently investigated in \cite{Ebadi:2023xhq} using a parameterised template for the spectator sourced curvature perturbation. In this work we study a concrete setup, the modulated reheating mechanism, defined at the level of the action.    

We investigate a modulated reheating setup realised through shift-symmetric dimension-five operators that couple a pseudo-scalar inflaton to fermion or vector fields whose masses depend on the spectator scalar field. The structure of the mass terms is similar to masses generated by the Higgs field in the Standard Model (SM) of particle physics. We assume the spectator scalar has a quartic self-interaction and a non-minimal coupling to spacetime curvature. The spectator sector has a Higgs-like structure and the setup could be realised in SM extensions that are nearly ultraviolet complete and can be perturbatively extrapolated up to the inflationary energy scale. Here we do not focus on any specific SM extension  but perform a phenomenological study treating the spectator couplings as free parameters.  

We require that the curvature perturbation on large scales $k \lesssim 0.1 {\rm Mpc}^{-1}$ is dominated by the inflaton sourced part $\zeta_{\phi}$ and model the inflaton sector by $R^2$ inflation \cite{Starobinsky:1979ty,Kehagias:2013mya}. We use the stochastic approach \cite{Starobinsky:1986fx,Starobinsky:1994bd} and the $\delta N $ formalism  \cite{Starobinsky:1985ibc,Salopek:1990jq,Sasaki:1995aw,Wands:2000dp,Lyth:2004gb} to compute the curvature perturbation $\zeta_{\chi}$ sourced by the spectator field and compute its correlators as ensemble expectation values in the de Sitter vacuum state. The de Sitter vacuum is the stationary asymptotic solution towards which the spectator distribution relaxes during de Sitter inflation with a rate proportional to the effective mass of the spectator field \cite{Starobinsky:1994bd, Enqvist:2012xn}. For the $R^2$ inflaton, the Hubble scale changes slowly in time and approximating the spectator distribution by the de Sitter equilibrium solution should be well justified. For general time-dependent inflationary backgrounds this may not always be the asymptotic solution, however \cite{Hardwick:2017fjo}. We also note that while long-wavelength fluctuations in general may generate differences between volume averages in the observed patch of the universe and ensemble expectation values, their impacts are suppressed in our setup where the effective spectator mass can be comparable to the Hubble scale, $m_{\rm eff}^2/H^2 \sim 0.1 $ \cite{Linde:2005yw}.  

The one-point function of the spectator field vanishes $\langle \chi\rangle = 0$ in the de Sitter vacuum and the curvature perturbation $\zeta_{\chi}$ sourced by it is completely non-Gaussian, having no leading Gaussian part \cite{Lyth:2001nq,Boubekeur:2005fj}.  Using the stochastic spectral expansion method \cite{Starobinsky:1994bd,Markkanen:2019kpv,Markkanen:2020bfc}, we compute the spectrum of the infrared two-point function of $\zeta_{\chi}$. We further estimate the bispectrum of the three-point function of $\zeta_{\chi}$ based on the exact result we find in the limit where the non-minimal coupling $\xi$ dominates the spectator dynamics during inflation and $\zeta_{\chi}$ is truncated to quadratic order in the spectator field. We scan over a wide range of the model parameter space, impose Planck constraints on the spectrum and non-Gaussianity, and compute the stochastic gravitational wave signal sourced by $\zeta_{\chi}$ at second order in perturbations for configurations where $\zeta_{\chi}$ dominates over $\zeta_{\phi}$ on small scales. 

The paper is organised as follows.  In Section 2 we present the setup, compute the curvature perturbation, and study bounds imposed by Planck constraints on non-Gaussianity. Details of the computations are presented in the Appendices. In Section 3 we briefly describe the computation of second order gravitational waves induced by the spectator fluctuations. In Section 4 we present our results, and finally conclude in Section 5. 

We use the $(-,+,+,+)$ signature for the metric and denote the reduced Planck mass by $M_{\rm P}\equiv (8\pi G)^{-1/2}$. 

\section{The setup}
\label{sec:section2}

We investigate a setup where primordial perturbations consist of a nearly scale-invariant Gaussian component sourced by the inflaton field $\phi$ and a blue-tilted non-Gaussian component sourced by a spectator scalar $\chi$. The spectator is energetically subdominant during inflation but it generates curvature perturbations via the modulated reheating mechanism  \cite{Kofman:2003nx,Dvali:2003em}. We assume a quartic self-interaction for the spectator and include its non-minimal coupling to the Ricci scalar, $\xi \chi^2 R$, in the Lagrangian. The non-minimal coupling is generated through one-loop radiative corrections in curved spacetime even if it would be absent in the tree-level action. As a concrete template for the inflaton sector we use the potential of the $R^2$ model of inflation \cite{Starobinsky:1979ty}.  

The relevant part of the Lagrangian in our setup is given by    
\beq
\label{eq:action}
{\cal L} = -\frac{1}{2}\nabla^{\mu}{\phi}\nabla_{\mu}{\phi}-\Lambda_{\phi}^4\left(1-e^{-\sqrt{2/3} \phi/M_{\rm P}}\right)^2-\frac{1}{2}\nabla^{\mu}\chi \nabla_{\nu} \chi-\frac{1}{2}\xi R \chi^2 -\frac{1}{4}\lambda \chi^4+{\cal L}_{\rm dec}(\phi,\chi,X)+\ldots, 
\eeq
where ${\cal L}_{\rm dec}(\phi,\chi,X)$ is specified below and gives rise to the modulated reheating mechanism such that the inflaton decays into thermal bath particles $X$ with a $\chi$-dependent decay rate. The ellipses denote all other terms, including $\chi$ dependent mass terms for $X$ particles that will be relevant in the setup and will be discussed below. In the following we will neglect all dynamical effects of ${\cal L}_{\rm dec}(\phi,\chi,X)$ during inflation. In other words, we parameterise the inflaton dynamics during inflation by the pure $R^2$ model and assume the subsequent reheating stage is described by an effective theory which involves ${\cal L}_{\rm dec}(\phi,\chi,X)$. 

We note that perturbations sourced by the modulated reheating mechanism depend on inflaton dynamics during reheating but are not directly affected by details of the inflaton potential in the inflationary region. Therefore, we expect that our results for the gravitational wave signal sourced by the modulated reheating mechanism are at least qualitatively representative for generic models where the inflaton potential during reheating takes an effectively quadratic form. On the other hand, our results cannot be carried over for models where the inflaton potential during reheating differs from the quadratic form and such setups should be studied separately.

We assume that during reheating the effective inflaton field is a gauge singlet pseudo-scalar and, following \cite{Lu:2019tjj}, we investigate two non-renormalisable dimension five templates for ${\cal L}_{\rm dec}(\phi,\chi,X)$ given by 
\baq
\label{eq:L1}
{\cal L}^{(1)}_{\rm dec}&=& -\frac{1}{\Lambda_{1}}\phi  F_{\mu\nu}\tilde{F}^{\mu\nu}~,\\
\label{eq:L2}
{\cal L}^{(2)}_{\rm dec}&=& -\frac{1}{\Lambda_{2}}{\bar{\psi}}  (\partial\hspace{-6pt}/ \phi) \gamma^{5}\psi~.
\eaq 
Here $F_{\mu\nu}$ ($\tilde{F}_{\mu\nu}$) denote the (dual) field strength tensors of a gauge field $A^{\mu}$, $\psi$ denotes a fermion, and $\Lambda_{1,2}$ are dimensional couplings associated to the non-renormalisable operators. Sums over internal degrees of freedom are implied. The axion like couplings (\ref{eq:L1}) and (\ref{eq:L2}) are shift symmetric, leaving the classical dynamics unchanged under $\phi \rightarrow \phi + c$ for constant $c$, and therefore would not spoil the flatness of the inflaton potential 
even if they would be present in the effective theory during inflation. They could however source non-Gaussian inflaton perturbations and enhance gravitational waves from vacuum fluctuations, see e.g. \cite{Barnaby:2011qe,Barnaby:2010vf,Cook:2011hg,Barnaby:2011vw,Meerburg:2012id}, and such contributions could be significant depending on the values of $\Lambda_{1,2}$. As stated above, we will not account for these effects but treat equations (\ref{eq:L1}) and (\ref{eq:L2}) as effective operators present only after the end of inflation. Our focus here is to study gravitational waves induced by the spectator field via the modulated reheating mechanism. As we will show in this work, this signal turns out to be unobservably small for all values of $\Lambda_{1,2}$ for which the modulation mechanism can be phenomenologically realised. Accounting for the dynamical effects of (\ref{eq:L1}) and (\ref{eq:L2}) during inflation could constrain the couplings $\Lambda_{1,2}$ cutting out part of our phenomenological parameter space but they would not affect our main finding of the unobservability of the gravitational wave signal. Moreover,  
while the modulated reheating dynamics is not directly affected by the inflaton potential in the inflationary regime, the constraints on operators (\ref{eq:L1}) and (\ref{eq:L2}) are strongly dependent on details of the inflaton model.

After the end of inflation, the inflaton field oscillates in an effectively quadratic potential with the mass $m_{\phi} = (2/\sqrt{3}) \Lambda_{\phi}^2/M_{\rm P}$. We assume the inflaton decay proceeds either through the operator (\ref{eq:L1}) or (\ref{eq:L2}), and refer to these as the vector and fermion setups, respectively. In the vector setup, we identify the spectator $\chi$ as the length of a scalar multiplet charged under the gauge group associated to $A^{\mu}$ and in the fermion setup we assume a Yukawa coupling between $\chi$ and $\psi$. The masses of $A^{\mu}$ and $\psi$ particles are then given by   
\beq
m_{A} = \frac{g}{2}\chi~, \qquad m_{\psi} = \frac{y_{\psi}}{\sqrt{2}}\chi ~,
\eeq
similar to the structure of the mass terms induced by the Higgs field in the SM. 

The decay rates for $\phi\rightarrow A A$ and $\phi\rightarrow \psi\bar{\psi}$ during the rapid inflaton oscillations in the quadratic potential can be approximated by \cite{Shtanov:1994ce,Ichikawa:2008ne,Lu:2019tjj,Garcia:2020wiy,Garcia:2023obw} 
\baq
\label{eq:gamma1}
\Gamma^{(1)}&=&\frac{m_{\phi}^3}{4\pi \Lambda_{1}^2}\left(1-\frac{g^2 \chi^2}{m_{\phi}^2}\right)^{3/2}~~~~\,\equiv ~\Gamma_0^{(1)}\left(1-\frac{g^2 \chi^2}{m_{\phi}^2}\right)^{3/2}~, \\
\label{eq:gamma2}
\Gamma^{(2)}&=&\frac{m_{\phi}m_{\psi_I}^2}{2\pi \Lambda_{2}^2}\left(1-\frac{2 y_{\psi}^2 \chi^2}{m_{\phi}^2}\right)^{1/2}\equiv ~\Gamma_0^{(2)}\left(1-\frac{2 y_{\psi}^2 \chi^2}{m_{\phi}^2}\right)^{1/2}~.
\eaq  
Here we have negleted multiplicative factors associated to the internal degrees of freedom of the final state particles.

Our setup assumes that the spectator does not decay before the inflaton has decayed into radiation and the universe has become radiation dominated. For definiteness we also assume that the spectator eventually decays into radiation and that the decay products thermalise with the existing radiation component so that no residual isocurvature perturbations are left. The curvature perturbations produced in the setup are not sensitive to details of the spectator decay, however, provided that it happens after the universe has become radiation dominated. In the radiation dominated universe $R=0$ and energy density of the spectator field oscillating in a quartic potential scales as radiation. Its contribution to the total energy density therefore remains constant, unlike for example in the curvaton scenario \cite{Lyth:2001nq,Moroi:2001ct,Enqvist:2001zp,Linde:1996gt,Mollerach:1989hu}. Since the spectator energy density is very small to start with (in our setup at the end of inflation $\rho_{\chi}/\rho_{\rm tot}\sim H^2/M_{\rm P}^2 \sim 10^{-10}$), the tiny isocurvature perturbations present after the inflaton decay until the decay of the spectator field have negligible effects on curvature perturbations.  

\subsection{Curvature perturbation}

Using the $\delta N$-formalism \cite{Starobinsky:1985ibc,Salopek:1990jq,Sasaki:1995aw,Wands:2000dp,Lyth:2004gb}, the curvature perturbation on superhorizon scales can be written as 
\beq
\label{eq:deltaNmain}
\zeta({\bf x}) = N(\phi({\bf x}),\chi({\bf x}))-\langle N(\phi({\bf x}),\chi({\bf x})) \rangle~.
\eeq
Here $N(\phi({\bf x}),\chi({\bf x})) = \int_{t_{\rm i}}^{t_{\rm f}} dt H({\bf x}) d t $ is the local number of e-folds from an initial uniform spatial curvature hypersurface at time $t_{\rm i}$ to a uniform energy density hypersurface at $t_{\rm f}$. We take the final time $t_{\rm f}$ to be after the inflaton has decayed so that perturbations have become adiabatic and the curvature perturbation is constant in time, $\zeta(t> t_{\rm f},{\bf x}) = \zeta(t_{\rm f},{\bf x}) \equiv\zeta({\bf x})$. The local number of e-folds is solved from equations of motion for a homogeneous and isotropic universe with the initial conditions $\phi({\bf x})\equiv \phi(t_{\rm i},{\bf x})$ and $\chi({\bf x})\equiv \chi(t_{\rm i},{\bf x})$. By construction, the choice of the initial time $t_{\rm i}$ does not affect the value of the curvature perturbation~\cite{Lyth:2004gb}.  

We neglect all contributions from the $\chi$ field in the Friedmann equations because we investigate only the parameter range where $\chi$ is an energetically subdominant spectator field during inflation and reheating. We further assume that the inflaton decay products thermalise instantaneously, forming an ultrarelativistic radiation component $\rho_{\rm r}$. 

The relevant equations of motion in our setup are given by 
\baq
\label{eq:eom}
\ddot{\phi}+3H\dot{\phi}+\Gamma(\chi)\dot{\phi}
+2\sqrt{\frac{2}{3}}\frac{\Lambda_{\phi}^4}{M_{\rm P}}\left(1-e^{-\sqrt{2/3} \phi/M_{\rm P}}\right)e^{-\sqrt{2/3} \phi/M_{\rm P}}&=&0~,\\
\nonumber
\ddot{\chi}+3H\dot{\chi}+\lambda\chi^3+\xi R \chi&=&0~,\\
\nonumber
\dot{\rho}_{\rm r}+4\rho_{\rm r}&=&\Gamma(\chi)\dot{\phi}^2~,\\
\nonumber
\frac{1}{3 M_{\rm P}^2}\left(\frac{1}{2}\dot{\phi}^2+\Lambda_{\phi}^4\left(1-e^{-\sqrt{2/3} \phi/M_{\rm P}}\right)^2+\rho_{\rm r}\right)&=&H^2~,
\eaq
where $\Gamma(\chi)$ is given by the real part of equation (\ref{eq:gamma1}) or (\ref{eq:gamma2}). The curvature scalar $R$ in the equation of motion for $\chi$ can be written as  
\beq
\label{eq:ricci}
R=3H^2-\frac{3}{M_{\rm P}^2}\left(\frac{1}{2}\dot{\phi}^2-\Lambda_{\phi}^4\left(1-e^{-\sqrt{2/3} \phi/M_{\rm P}}\right)^2+\frac{1}{3}\rho_{\rm r}\right)~.
\eeq 

We choose the initial time $t_{\rm i}$ slightly before the end inflation such that $\epsilon_{\rm H}(t_{\rm i}) =0.1$, where $\epsilon_{\rm H} = -\dot{H}/H^2$, and set the initial conditions for equations (\ref{eq:eom}) as follows:  For the radiation component we set $\rho_{\rm r}(t_{\rm i}) = 0$. For the spectator field we set $\dot{\chi}(t_{\rm i}, {\bf x}) = 0$ and draw the field value $\chi(t_{\rm i}, {\bf x})\equiv \chi({\bf x})$ from the de Sitter equilibrium distribution given by \cite{Starobinsky:1994bd}  
\beq
\label{eq:psi02}
P(\chi) = C_0^{-1} e^{-\frac{8\pi^2}{3 H_{\rm i}^4}\left(\frac{\lambda}{4}\chi^4+6\xi H_{\rm i}^2\chi^2 \right)}~,\qquad C_0 = \int_{-\infty}^{\infty} d\chi e^{-\frac{8\pi^2}{3 H_{\rm i}^4}\left(\frac{\lambda}{4}\chi^4+6\xi H_{\rm i}^2\chi^2 \right)}~.
\eeq
For the inflaton sector, we first solve for the pure inflaton system, corresponding to (\ref{eq:eom}) with $\Gamma = 0, ~\rho_{\rm r}= 0, ~\chi = 0$. For this system we set slow roll initial conditions $\phi_0 = 5.8 M_{\rm P}$ and $\dot{\phi}_0= -(2\sqrt{2}/3) \Lambda_{\phi}^2 {\rm exp}(-\sqrt{2/3}\phi_0/M_{\rm P})$ which correspond to an initial time $N(t_0)\gg 60$ e-folds before the end of inflation. For the full system (\ref{eq:eom}), we then set $\phi(t_{\rm i},{\bf x}) = \bar{\phi}(t_{\rm i})+ \delta\phi(t_{\rm i},{\bf x})\equiv \bar{\phi}+ \delta\phi({\bf x}) $ where $\bar{\phi}(t_{\rm i})$ is the homogeneous solution of the pure inflaton system at $\epsilon_{\rm H}=0.1$  and $\delta\phi(t_{\rm i},{\bf x})$ is the standard linear perturbation theory result for the inflaton perturbation in the pure inflaton system in the spatially flat gauge at  $\epsilon_{\rm H}=0.1$. 

We evolve the system of equations (\ref{eq:eom}) until a final time $t_{\rm f}$ defined as $\Omega_{\phi}(t_{\rm f}) \equiv \rho_{\phi}(t_{\rm f})/\rho_{\rm tot}(t_{\rm f})= 10^{-4}$ well after the reheating has completed. We have checked that $\zeta$ has settled to a constant value by our $t_{\rm f}$, and defining $t_{\rm f}$ as a later time event does not change our results. 

Writing $\phi({\bf x}) = \bar{\phi}+ \delta\phi({\bf x})$, expanding to first order in $\delta\phi$, and using that  $\langle\delta \phi\rangle = 0 $, we can recast equation  (\ref{eq:deltaNmain}) as 
\beq
\label{eq:zetafull}
\zeta({\bf x}) = \zeta_{\phi}({\bf x})+\zeta_{\chi}({\bf x})~,
\eeq
where
\baq
\label{eq:zetaphi}
\zeta_{\phi}({\bf x})  &=& \partial_{\phi}N(\bar{\phi},\chi({\bf x})) \delta\phi({\bf x})\\
\label{eq:zetachi}
\zeta_{\chi}({\bf x})  &=& N(\bar{\phi},\chi({\bf x}))-\langle N(\bar{\phi},\chi({\bf x})) \rangle~.
\eaq
Our setup does not involve direct non-gravitational couplings between $\phi$ and $\chi$. Neglecting also the slow-roll suppressed correlators generated via the non-minimal coupling, see e.g. \cite{Markkanen:2017dlc}, we have $\langle \delta\phi \chi^n\rangle = 0$ for all values of $n$ and $\langle \zeta_{\phi}\zeta_{\chi}\rangle =0$. The power spectrum of the two-point function $\langle \zeta({\bf k})\zeta({\bf k'})\rangle = (2\pi)^3\delta({\bf k}+{\bf k}') (2\pi^2/k^3) {\cal P}_{\zeta}(k)$ is then given by the sum
\beq
\label{eq:Pzetasum}
{\cal P}_{\zeta}(k) = {\cal P}_{\zeta_{\phi}}(k) + {\cal P}_{\zeta_{\chi}}(k)~,
\eeq
where ${\cal P}_{\zeta_{\phi}}$ and ${\cal P}_{\zeta_{\chi}}$ denote the spectra of $\zeta_{\phi}$ and $\zeta_{\chi}$, respectively. 

The $\chi({\bf x})$ dependence drops out from $\partial_{\phi}N$ in equation (\ref{eq:zetaphi}), see Appendix \ref{app:A} for details. The power spectrum of  the inflaton sourced curvature perturbation $\zeta_{\phi}$ is then given by the standard expression 
\beq
\label{eq:Pzetaphi}
{\cal P}_{\zeta_{\phi}}(k)= \frac{M_{\rm P}^{-2}}{2\epsilon(t_k)}\left(\frac{H(t_k)}{2\pi}\right)^2~ ,
\eeq
where $t_{k}$ is defined via $k=a(t_k)H(t_k)$ and computed using the background solution $\chi=0$. 

In order to determine the power spectrum of the spectator sourced part $\zeta_{\chi}$, we apply the stochastic formalism \cite{Starobinsky:1986fx,Starobinsky:1994bd} and the spectral expansion method \cite{Starobinsky:1994bd,Markkanen:2019kpv,Markkanen:2020bfc}. Here we neglect the time dependence of the Hubble rate and perform the computation using de Sitter results with $H=H(t_{\rm i})$. Moreover, we assume that the distribution for $\chi(\bf{x})$ has relaxed to the de Sitter equilibrium state. In particular, the probability distribution for the one-point function is then given by equation (\ref{eq:psi02}). Under these assumptions the power spectrum ${\cal P}_{\zeta_{\chi}}$ takes the form 
\beq
\label{eq:Pzetachi_sum}
{\cal P}_{\zeta_{\chi}}(k)=\sum_{n=0}^{\infty} \left(\int\hspace{-2pt}{\rm d} \chi \psi_0(\chi)\zeta_{\chi}(\chi)\psi_n(\chi)\hspace{-2pt}\right)^{\hspace{-4pt}~2} 
\hspace{-2pt}\frac{2}{\pi} \Gamma\hspace{-2pt}\left(\hspace{-2pt}2-\frac{2\Lambda_n}{H}\hspace{-2pt}\right){\rm sin}\hspace{-2pt}\left(\hspace{-2pt}\frac{\Lambda_n}{H}\hspace{-2pt}\right)
\hspace{-2pt}
\left(\hspace{-2pt}\frac{k}{aH}\hspace{-2pt}\right)^{\hspace{-2pt}\frac{2\Lambda_n}{H}}\hspace{-8pt}\theta(k_{\rm cut}-k)~.
\eeq
Details of the computation are given in Appendix \ref{app:A}. The eigenfunctions $\psi_n(\chi)$ and eigenvalues $\Lambda_n$ are determined by the equation 
\beq
\psi_n''(\chi) + \left(U''(\chi)-U'(\chi)\right) \psi_n(\chi)= -\frac{4\pi^2 \Lambda_n}{H^3}\psi_{n}(\chi)~,
\eeq
with the asymptotic boundary conditions, $\psi_n(\chi) \rightarrow 0 $ for $\chi\rightarrow \pm\infty$, and
\beq
\label{eq:Uchi}
U(\chi)=\frac{4\pi^2}{3H^4}\left(\frac{1}{2}\left(m^2+12\xi H^2\right)\chi^2 +\frac{1}{4}\lambda\chi^4\right)~.
\eeq 

The ultraviolet cutoff $k_{\rm cut}\equiv a(t_{\rm f})H(t_{\rm f})$ in equation (\ref{eq:Pzetachi_sum}) is introduced because the $\delta N$ approach holds only for superhorizon modes. We solve equations (\ref{eq:eom}) up to the final time $t_{\rm f}$ and our results therefore apply for modes $k<a(t_{\rm f})H(t_{\rm f})$. As we will discuss in Section \ref{sec:results}, the cutoff scale $k_{\rm cut}$ implied by our definition of $t_{\rm f}$ via $\Omega_{\phi}(t_{\rm f}) = 10^{-4}$ is much above the scales relevant for gravitational wave interferometer surveys. 

We find that the sum (\ref{eq:Pzetachi_sum}) is very well approximated by its first non-vanishing term over all scales relevant in our analysis and in the coupling range where the spectrum is compatible with existing observational bounds.
In the following we will therefore compute ${\cal P}_{\zeta_{\chi}}$ using  
\beq
\label{eq:Pzetachi}
{\cal P}_{\zeta_{\chi}}(k)= 
\left(\int\hspace{-2pt}{\rm d} \chi \psi_0(\chi)\zeta_{\chi}(\chi)\psi_2(\chi)\hspace{-2pt}\right)^{\hspace{-4pt}~2} 
\hspace{-2pt}\frac{2}{\pi} \Gamma\hspace{-2pt}\left(\hspace{-2pt}2-\frac{2\Lambda_2}{H}\hspace{-2pt}\right){\rm sin}\hspace{-2pt}\left(\hspace{-2pt}\frac{\Lambda_2}{H}\hspace{-2pt}\right)
\hspace{-2pt}
\left(\hspace{-2pt}\frac{k}{aH}\hspace{-2pt}\right)^{\hspace{-2pt}\frac{2\Lambda_2}{H}}\hspace{-8pt}\theta(k_{\rm cut}-k)~.
\eeq

\subsection{CMB constraints}

The amplitude and the spectral index ($n_{\rm s}-1 \equiv {\rm d}\,{\rm ln}\, {\cal P}_{\zeta}/{\rm d}\,{\rm ln}\, k $) of primordial perturbations measured by Planck are given by \cite{Planck:2018vyg}
\beq
\label{eq:planckspectrum}
{\cal P}_{\zeta}(k_{*}) = (2.100 \pm 0.030)\times 10^{-9}~,\qquad n_{\rm s}(k_{*})=0.9649\pm0.0042~,
\eeq
at the CMB pivot scale $k_{*} = 0.05\, {\rm Mpc}^{-1}$.
We require that ${\cal P}_{\zeta}(k_{*})$ and $n_{\rm s}(k_{*})$ computed from equation (\ref{eq:Pzetasum}) lie within these $1 \sigma$ intervals. When the spectator contribution in our setup is relevant, it also generates a positive running of the spectral index  
$\alpha_{\rm s}\equiv  {\rm d}\,n_{\rm s}/{\rm d}\,{\rm ln}\, k>0$. The Planck bound~\cite{Planck:2018vyg} is $\alpha_{\rm s}=-0.0045\pm 0.0067$ \cite{Planck:2018vyg} so we further require that the running computed from (\ref{eq:Pzetasum}) obeys $\alpha_{\rm s}(k_*) < 0.0022$. 

The latest data release of the Atacama Cosmology Telescope (ACT) reports a higher value for the spectral index $n_{\rm s}(k_{*})=0.974\pm0.003$ \cite{ACT:2025fju} compared to Planck. The dominant CMB constraints for our setup come from the non-Gaussianity discussed below. Using the ACT data instead of the Planck data would therefore not change our main results on spectator sourced gravitational waves but one would have to model the inflaton sector with something else than the $R^2$ inflation to get a good fit for the spectral index. 

Observational bounds on non-Gaussianity on the CMB scales \cite{Planck:2019kim} place stringent constraints for the spectator contribution in our setup. For the decay rates given by equations (\ref{eq:gamma1}) and (\ref{eq:gamma2}), $N(\phi,\chi)$ is an even function of $\chi$, and $\zeta$ contains no linear term in $\chi$. Moreover, we compute ensemble expectation values in the Sitter equilibrium where there is no classical spectator field background $\langle \chi \rangle = 0$. From this it follows that $\zeta_{\chi}$ contains no leading Gaussian term but it is a manifestly non-Gaussian component. Therefore,   $\zeta_{\chi}$ has to be sufficiently subdominant compared to $\zeta_{\phi}$ on large scales, $k \gtrsim 0.1 {\rm Mpc}^{-1}$ \cite{Boubekeur:2005fj}.    

In Appendix \ref{app:B}, we compute the full momentum-dependent bispectrum of $\zeta_{\chi}$ under the approximation that equation (\ref{eq:zetachi}) is truncated to the leading (quadratic) order in the expansion around $\chi = 0$. The computation also assumes that $\chi$ is a Gaussian field which in our setup turns out to be a good approximation for $\lambda \lesssim 10^{-3}$ when $\xi \gtrsim 0.01$. The bispectrum obtained in Appendix \ref{app:B} is of nearly local shape so we can apply the Planck constraint for the local non-Gaussianity parameter $f^{\rm local}_{\rm NL} = -0.9 \pm 5.1$ \cite{Planck:2019kim}. Requiring the maximal value of the weakly scale-dependent positive non-Gaussianity parameter obtained in Appendix \ref{app:B} to be below the Planck bound, $f_{\rm NL}^{\rm max} < 4.2$, in the entire coupling range $\xi < 0.04$ studied in Section \ref{sec:results}, we obtain the constraint  ${\cal P}_{\zeta_{\chi}}(k_*) < 5\times 10^{-12}$ for the spectrum of the spectator sourced component at the CMB pivot scale. 

The error made by truncating equation (\ref{eq:zetachi}) in Appendix \ref{app:B} depends on values of $\lambda$ and $\xi$, as well as $g$ or $y_{\psi}$ in the decay rate (\ref{eq:gamma1}) or (\ref{eq:gamma2}), respectively. Using the stochastic formalism we have compared the contact limit three-point functions $\langle \zeta_{\chi}(\bf{x})\zeta_{\chi}(\bf{x})\zeta_{\chi}(\bf{x}) \rangle$ computed using the full equation (\ref{eq:zetachi}) for $\zeta_{\chi}$ and its truncation (\ref{eq:appzeta}). For $\lambda \lesssim 10^{-4}$, the truncation gives an ${\cal O}(1)$ estimate for the full contact limit three-point function when ${\cal P}_{\zeta_{\chi}}(k_*) < 5\times 10^{-12}$. We therefore expect $f^{\rm max}_{\rm NL}$ to give an ${\cal O}(1)$ estimate for the full non-linearity parameter in this region. Requiring that $f^{\rm max}_{\rm NL}$ computed in the Appendix \ref{app:B} is conservatively at least factor $5$ below the Planck $1\sigma$ bound $f^{\rm local}_{\rm NL} = -0.9 \pm 5.1$ in the entire range $\xi < 0.04$ sets the bound 
\beq
\label{eq:pchibound}
{\cal P}_{\zeta_{\chi}}(k_*) < 10^{-12}~.
\eeq 
For $\lambda \sim 10^{-3}$, the contact limit three point function computed using the truncation (\ref{eq:appzeta}) starts to deviate from the full result by a factor up to ${\cal O}(10)$ for very large couplings $g, y_{\psi}\gtrsim 4$ but the truncation always overestimates the full result. The same seems to hold true for $\lambda \gtrsim 10^{-2}$ where the eigenfunctions also start to deviate considerably from the $\lambda = 0$ solutions. The inequality (\ref{eq:pchibound}) should therefore give a conservative constraint sufficient to comply with the Planck bounds in the entire parameter range studied in this work and we will apply it in this sense in our analysis.  

In addition to the bispectrum, the Planck data also constrains the trispectrum, i.e. the four-point function of $\zeta$. Approximating $\zeta_{\chi}$ by the truncated form in Appendix \ref{app:B}, one obtains a parameteric estimate for the trispectrum non-linearity parameter $g_{\rm NL} \sim {\cal P_{\zeta_{\chi}}}^2/{\cal P_{\zeta}}^3$. Using the bound (\ref{eq:pchibound}), this gives  $g_{\rm NL} \lesssim 10^2$ which is well below the Planck constraint for the local trispectrum amplitude, $g_{\rm NL}^{\rm local} = (-5.8\pm 6.5)\times 10^4$ \cite{Planck:2019kim}.

\section{Computation of the stochastic gravitational wave spectrum}
\label{sec:SIGW}

Scalar induced gravitational waves emerge as second-order tensor perturbations sourced by first-order scalar perturbations~\cite{Ananda:2006af,Baumann:2007zm}. The gravitational wave signal sourced by the spectator perturbations in our setup can be straightforwardly computed applying well-known results from the literature~\cite{Ananda:2006af,Baumann:2007zm,Kohri:2018awv}. Here we briefly present the relevant steps of the computation. 
  
We focus on modes $k < k_{\rm cut}$ which enter the horizon when the universe in our setup has become radiation dominated and perturbations are adiabatic on superhorizon scales. In the radiation dominated universe, 
 the second order tensor perturbations follow the equation of motion
    \begin{equation}
        h_{\lambda}^{\prime\prime}(\tau,\veck) + \frac{2}{\tau} h_{\lambda}^{\prime}(\tau,\veck) + k^2 h_{\lambda}(\tau,\veck) = 4 \mathcal{S}_\lambda(\tau,\veck)\;,
        \label{eq:EOM_h}
    \end{equation}
where $\lambda$ is the polarization index, $\tau$ is the conformal time and prime denotes the derivative with respect to $\tau$.  The source term, $\mathcal{S}_\lambda(\tau,\veck)$ is given by~\cite{Baumann:2007zm,Ananda:2006af}
    \begin{eqnarray}
        \mathcal{S}_\lambda(\tau,\veck) = 
        \int \dfrac{d^3 \bvec q}{(2\pi)^3} \, 
        \epsilon^{ij}_{\lambda}(\veck) q_i q_j
        \lrBigsb{
        & 3 \Phi_{\bvec q}\Phi_{{\bf k-q}}+
        2 \tau \Phi_{\bvec q}^{\prime}\Phi_{\bf k-q}+
         \tau^2 \Phi_{\bf k-q}^{\prime}\Phi_{\bvec q}^{\prime}
              } \;,
        \label{eq:source}
    \end{eqnarray}
where $\epsilon^{ij}_{\lambda}(\veck)$ is the polarization tensor ($\epsilon^{ij} k_i = \epsilon^{i}_{~i} = 0 $, $\sum_{\lambda}e^{ij}_{\lambda}e_{ij,\lambda} = 2$), and the Bardeen potential $\Phi$ is defined in the Newtonian gauge as  $ds^2 = a^2(-(1+2\Phi) \, d\tau^2 + (1-2\Phi) \, \delta_{ij} \, dx^i dx^j)$. 

The Bardeen potential in the radiation dominated universe is given by~\cite{Baumann:2007zm,Ananda:2006af}  
   \begin{equation}
        \Phi_{\bf k}(\tau) = \dfrac{2}{\lrb{k\tau/\sqrt{3}}^3} 
            \lrb{ \sin \dfrac{k\tau}{\sqrt{3}} -  \dfrac{k\tau}{\sqrt{3}} \cos \dfrac{k\tau}{\sqrt{3}}}  \zeta(k)~,
        \label{eq:Phi_sol}
    \end{equation}
where $\zeta(k)$ is the constant superhorizon curvature perturbation given by the Fourier transform of equation \eqref{eq:Pzetachi}. Here it is understood that we restrict ourselves to the parameter regime where the spectator component dominates the curvature perturbation $\zeta(k) \approx \zeta_{\chi}(k)$ on scales under consideration. The solution to equation \eqref{eq:EOM_h} is obtained by the Green's function method as  
    \begin{equation}
        h_{\lambda}(\tau,\veck) =
        \dfrac{4}{\tau}\int^\tau d\tilde{\tau} \tilde{\tau}\mathcal{S}_\lambda(\tilde{\tau},\veck) \dfrac{\sin k (\tau-\tilde{\tau}) }{k} ~.
        \label{eq:h_sol_G}
    \end{equation}

We define the spectrum of gravitational waves in the usual manner as
    \begin{equation}
        \svev{ h_{\lambda}(\tau,\veck) h_{\lambda'}(\tau,\bvec k') } = 
        (2\pi)^3  \delta(\bvec k + {\bf k}')\delta_{\lambda \lambda'} \frac{2\pi^2}{k^3}{\cal P}_h(\tau,k)~.
           \label{eq:h_correlator}
  \end{equation}
The two-point correlators of both polarization modes have equal amplitudes because, to the precision of our computation,  $\zeta_{\chi}({\bf k})$ does not acquire parity violating components from the operators (\ref{eq:L1}) and (\ref{eq:L2}), and perturbations are therefore symmetric under ${\bf k}\rightarrow -{\bf k}$. The gravitational wave signal is customarily characterised in terms of the spectral density fraction 
    \begin{equation}
         \Omega_{\rm GW}(\tau,k)  = 
         \dfrac{1}{3 H(\tau)^2 M_{\rm P}^2 } \dfrac{d \rho_{\rm GW}(\tau)}{d \log k} =  \dfrac{1}{24} \lrb{\dfrac{k}{a H}}^2  
       {\cal P}_h(\tau,k)~,
        \label{eq:Omega_GW_def}
    \end{equation}
where $\rho_{\rm GW}$ is the gravitational wave energy density, the second equality holds on subhorizon scales $k \gg a H$, and ${\cal P}_h(\tau,k)$ is the spectrum averaged over the oscillation cycle, see e.g. \cite{Domenech:2021ztg}. 

The gravitational wave spectrum ${\cal P}_h(\tau,k)$ computed using equations \eqref{eq:source}, \eqref{eq:Phi_sol} and \eqref{eq:h_sol_G},   contains a convolution integral over the four-point function $\langle \zeta({\bf q}) \zeta({\bf k}-{\bf q}) \zeta({\bf q}') \zeta({-\bf k}-{\bf q}')\rangle$ with ${\bf q}$ and ${\bf q}'$ being the integration variables. It has been shown that the gravitational wave component sourced by the disconnected part of the four-point function dominates over the contribution sourced by the connected part if $\zeta$ can be well approximated by an expansion around a leading Gaussian part \cite{Garcia-Saenz:2022tzu}. The connected part may however be important in strongly non-Gausssian setups, see e.g. \cite{Iovino:2024sgs, Zeng:2025cer}. Our setup is strongly non-Gaussian because $\zeta$ on small scales is dominated by $\zeta_{\chi}$ which contains no leading Gaussian term. Therefore, there is no a priori reason to neglect the connected part in our setup. Indeed, using the stochastic formalism we have checked that  the connected and disconnected parts of the coordinate space four-point function, $\langle \zeta_{\chi}({\bf x}_1) \zeta_{\chi}({\bf x}_2)\zeta_{\chi}({\bf x}_3) \zeta_{\chi}({\bf x}_4)\rangle$, are of the same order of magnitude in the contact limit $|{\bf x}_i- {\bf x}_j| \rightarrow 0 $. Computing the connected four-point function for general momentum configurations is however beyond the scope of this work. In the following, we will therefore investigate only the gravitational wave signal sourced by the disconnected part which is expected to give an order of magnitude estimate for the full gravitational wave signal.  

Discarding the connected part, the result for $\Omega_{\rm GW}(\tau,k)$ obtained in \cite{Ananda:2006af,Baumann:2007zm} can be written as
 \beq
         \Omega_{\rm GW}(k)  =  \dfrac{1}{24} \int_{0}^{\infty} {\rm d}t \int_{-1}^{1} {\rm d}s 
         \left(\frac{t(1-s)(2+t)}{(1+t+s)(1+t-s)}\right)^2
        {\cal P}_{\zeta}(u k) {\cal P}_{\zeta}(v k) I^2(u,v)\;,   
        \label{eq:Omega_GW_P_zeta_expression}
\eeq
where $u=\frac{1}{2}(1+s+t)$, $v=\frac{1}{2}(1-s+t)$, and $I^2(u,v) = I^2_A(u,v)(I_B^2(u,v) + I_C^2(u,v))$ with 
   \begin{align}
        I_A(u,v) &= \dfrac{3}{4}\dfrac{u^2+v^2-3}{u^3 v^3}~,
        \label{eq:Kd_factors}\\
        I_B(u,v) &= -4 \ u \, v \ + \ (u^2+v^2-3) \, \log \left| \dfrac{3-(u+v)^2}{3-(u-v)^2} \right|~,
        \\
        I_C(u,v) &= \pi \ (u^2+v^2-3) \theta(u+v-\sqrt 3) ~.
    \end{align}
The result for $\Omega_{\rm GW}(k)$ is constant in time during radiation domination as $\rho_{\rm GW}\propto a^{-4}$ and changes in the effective number of relativistic species $g_{*}(T)$ were neglected in the computation. Later changes in $g_{*}(T)$ can be accounted for by evaluating equation (\ref{eq:Omega_GW_P_zeta_expression}) at a reference time $\tau_{\rm c}$ sufficiently long after the horizon crossing of modes relevant for the gravitational wave spectrum, and then using the full time-dependent expression $\rho_{\rm tot}= 3 H^2 M_{\rm P}^2$ together with $\rho_{\rm GW}\propto a^{-4}$ for $\tau> \tau_{\rm c}$ . The result for the gravitational wave spectral density fraction in the universe today is given by \cite{Domenech:2021ztg}  
\beq
\label{eq:Omega_GW_0}
\Omega_{{\rm GW},0}(k)h^2 = \Omega_{{\rm r},0}h^2 \left(\frac{g_{*,{\rm c}}}{g_{*,0}}\right)\left(\frac{g_{*{\rm s},{\rm c}}}{g_{*{\rm s},0}}\right)^{-4/3}\Omega_{\rm GW}(k)~, 
\eeq 
where $\Omega_{{\rm r},0} = 4.18 \times 10^{-5} h^{-2}$ is the radiation density fraction in the present universe \cite{Planck:2018vyg} and $h\equiv H_0/(100 {\text{ km}}/{\text{s}}/{\text{Mpc}})$.  Modes with $k \gtrsim 10^{13}{\text{ Mpc}}^{-1}$ enter the horizon before the electroweak transition and, as can be seen from the results shown in the next section, this is the interesting range in our setup in light of the sensitivity curves of foreseeable gravitational wave surveys.  Focusing on modes in this range, we set $g_{*,{\rm c}}=g_{*{\rm s},{\rm c}}=106.75$ which yields $\Omega_{{\rm GW},0}(k)h^2 \simeq 1.6\times 10^{-5} \Omega_{\rm GW}(k)$. 

In order to compute $\Omega_{{\rm GW},0}(k)h^2$, we numerically evaluate the integrals in equation (\ref{eq:Omega_GW_P_zeta_expression}). For a power-law spectrum, such as in our case on scales relevant for the gravitational wave signal, this was done already in~\cite{Kohri:2018awv} and recently discussed in further detail in~\cite{Ebadi:2023xhq}. Our results are compatible with these works. In Appendix \ref{app:numericsGW} we briefly describe our method for dealing with the integrable singularity encountered at  $t=\sqrt 3 - 1$.

\section{Results}
\label{sec:results}

In this section we present our results for the gravitational wave density fraction (\ref{eq:Omega_GW_0}) produced by spectator sourced perturbations.  

Because of the non-Gaussianity bound (\ref{eq:pchibound}), the curvature perturbation on large scales must be dominated by the inflaton contribution (\ref{eq:Pzetaphi}). The measured Planck values for the power spectrum and spectral tilt (\ref{eq:planckspectrum}) set $\Lambda_\phi \approx 8\times 10^{15}~\GeV$, as usual in the $R^2$ inflation, with the exact value depending on the spectator couplings that affect the reheating history.  This fixes the Hubble scale in equation (\ref{eq:Pzetachi}) to $H\approx 1.16 \times 10^{13}~{\rm GeV}$. 

The slope of the spectator spectrum (\ref{eq:Pzetachi}) is determined by the couplings $\xi$ and $\lambda$ which also affect its amplitude. We investigate the range $10^{-3} < \xi < 10^{-1} $ and $10^{-6} < \lambda < 10^{-2}$ which spans spectator spectra from nearly scale-invariant to strongly blue-tilted ones. The amplitude of the spectrum depends further on the couplings $g$ and $\Lambda_{1}$, or $y_{\psi}$ and $\Lambda_{2}$, associated to the vector (\ref{eq:gamma1}) and fermion (\ref{eq:gamma2}) decay channels respectively. We study both channels separately and vary the respective couplings in the range $0.1<g<8$ and $0.1<\sqrt{2} y_{\psi}<8$. In both cases we set $\Gamma_0^{(1,2)}/m_{\phi} = 0.5^2/(8\pi) $ which corresponds to $\Lambda_{(1)} \approx 2.8 m_{\phi}$ and $\Lambda_{(2)} = 4 m_{\psi}$, respectively. 
Large coupling values, $g,y_{\psi}\gg1$, would arguably be difficult to realise within nearly ultraviolet complete SM extensions extrapolated to the inflationary scale. 
Moreover, even if such extensions could be found, the large couplings might be close to Landau poles indicating even more fundamental problems, or at least calling for non-perturbative methods in the analysis. 
Our results in the region $g,y_{\psi}\gg1$ should therefore be primarily understood in a phenomenological sense, showing how large the gravitational wave amplitude can maximally be assuming the inflation decay rate can be parameterised by equations (\ref{eq:gamma1}) and (\ref{eq:gamma2}).
For the fermion channel, we are also pushing to the limits of the effective field theory approach with the non-renormalisable operator (\ref{eq:L2}) since $\sqrt{\langle m_{\psi}^2\rangle}\sim \Lambda_{(2)}$ can be up to two orders of magnitude below the characteristic energy scale of the system, $m_{\phi}$.    

\begin{figure}[t!]
	\centering
	\begin{subfigure}[b]{0.47\textwidth}
		\includegraphics[width=\linewidth]{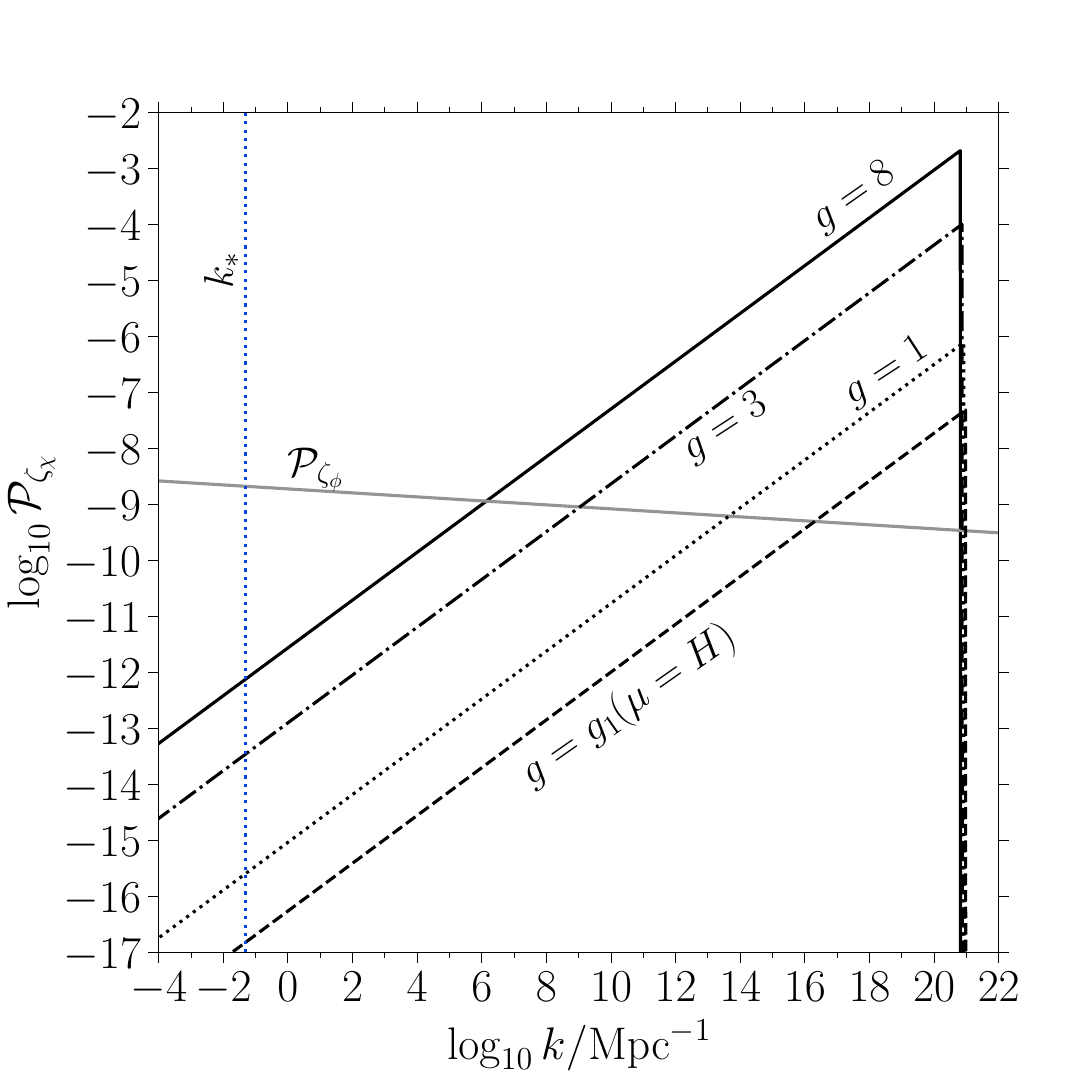}
		\label{fig:vector_spectrum_curves}
	\end{subfigure}
	\hfill
	\begin{subfigure}[b]{0.47\textwidth}
		\includegraphics[width=\linewidth]{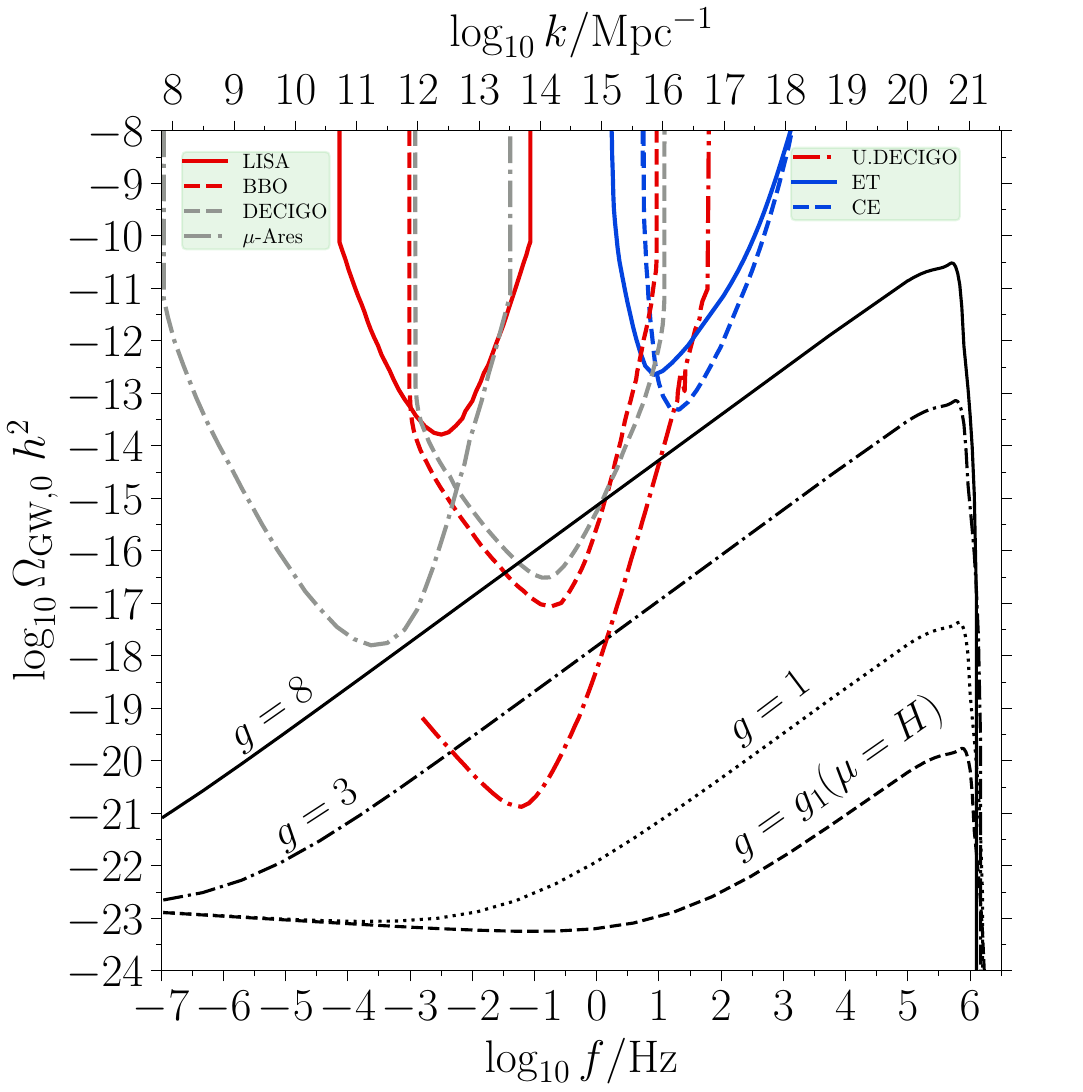}
		\label{fig:vector_curves}
	\end{subfigure}
	\centering
	\begin{subfigure}[b]{0.47\textwidth}
		\includegraphics[width=\linewidth]{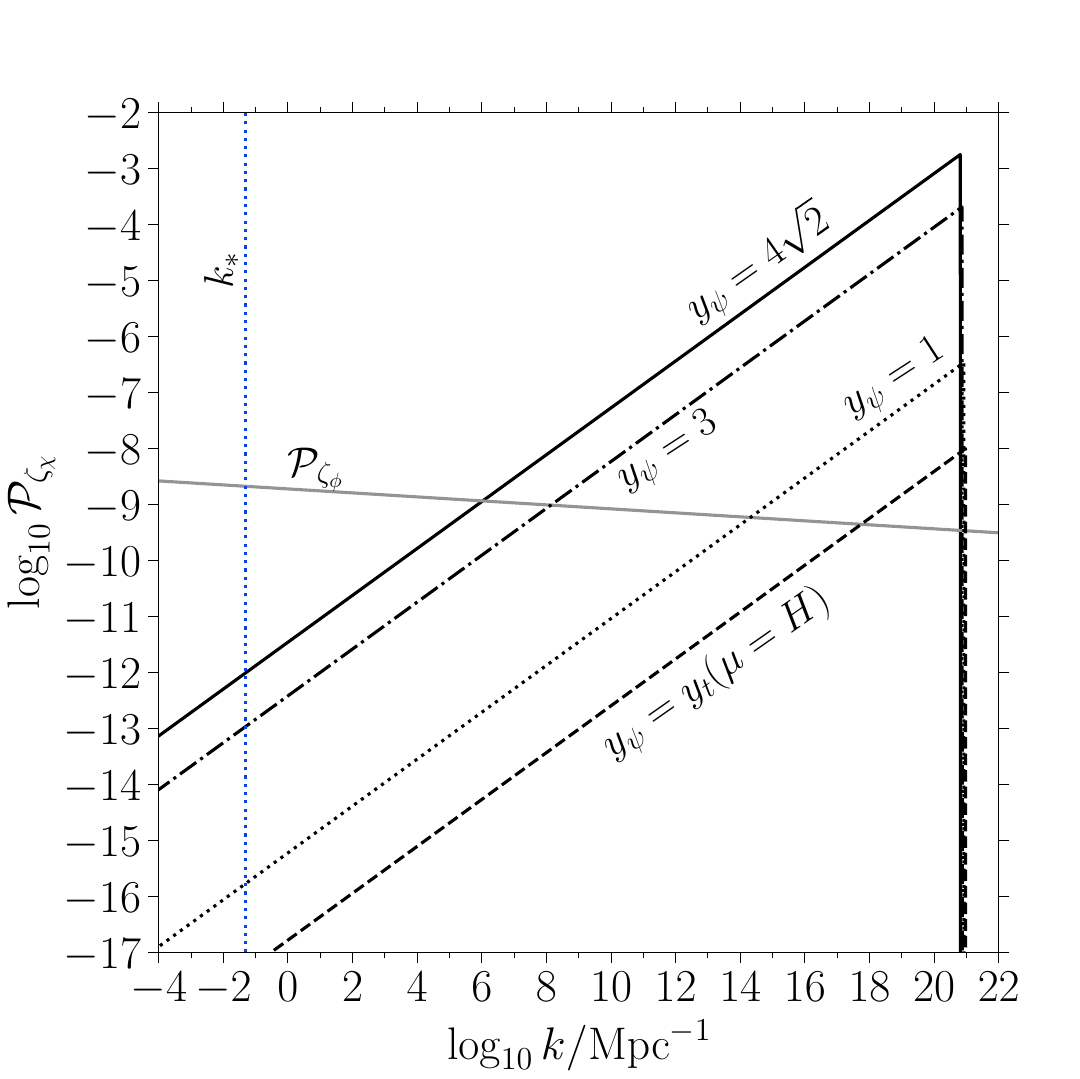}
		\label{fig:fermion_spectrum_curves}
	\end{subfigure}
	\hfill
	\begin{subfigure}[b]{0.47\textwidth}
		\includegraphics[width=\linewidth]{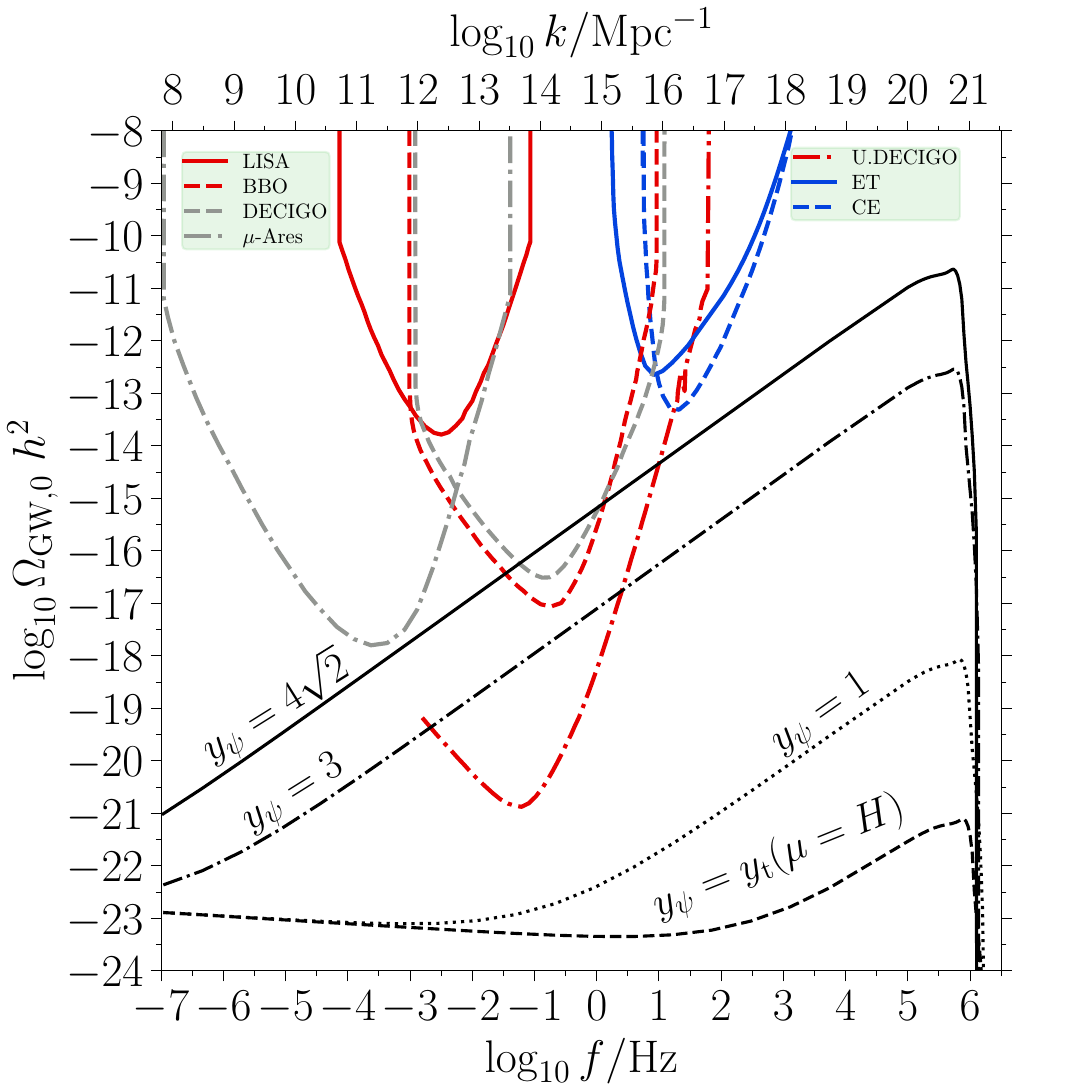}
		\label{fig:fermion_curves}
	\end{subfigure}
	\caption{The spectator sourced power spectrum ${\cal P}_{\zeta_\chi}$ (left panels) and the gravitational wave density fraction today $\Omega_{\rm GW,0}$ sourced by it (right panels). Upper (lower) panels show results for the vector (fermion) decay channel for different values of the coupling $g$ $(y_{\psi})$ and $\lambda  = 10^{-5}$. In the upper (lower) panels $\xi=0.0266$ ($\xi = 0.0262$). The dashed lines (lowest curves) show the results for coupling values equal to the SM weak gauge coupling $g_1$ and the top Yukawa coupling $y_{\rm t}$ evaluated at $\mu = H$. The grey lines in the left panels show the inflaton sourced power spectrum ${\cal P}_{\zeta_\phi}$. The right panels also show the power-law integrated sensitivity curves with SNR $=1$ for $\mu$-Ares (grey dashed-dotted), BBO (red dashed), LISA (red solid), DECIGO (grey dashed), ET (blue), CE (blue dashed), and Ultimate DECIGO (red dashed-dotted). 
	}
	\label{fig:curves}
\end{figure}

Figure~\ref{fig:curves} shows examples of the spectator sourced power spectrum ${\cal P}_{\zeta_{\chi}}(k)$ and the corresponding present day gravitational wave density fraction $\Omega_{\rm GW,0}(k)$ together with the sensitivity curves of future gravitational wave surveys  LISA \cite{2017arXiv170200786A,LISA:2024hlh}, BBO \cite{Crowder:2005nr,Harry:2006fi}, DECIGO and Ultimate DECIGO \cite{Seto:2001qf,Kawamura:2006up,Kudoh:2005as},  $\mu$-Ares \cite{Sesana:2019vho},  ET \cite{Punturo:2010zz,ET:2019dnz} and CE \cite{Reitze:2019iox,Evans:2021gyd}.
The depicted lines for each survey are the power-law integrated sensitivity curves for signal-to-noise ratio SNR $= 1$ and are taken from ~\cite{Sesana:2019vho,Schmitz:2020syl,Braglia:2021fxn}.  
Upper panels show results for the vector channel for different values of $g$ with $\xi=0.0266$ and $\lambda=10^{-5}$, including the case with $g$ equal to the SM SU(2) gauge coupling $g_1=\sqrt{5/3} g'$ evaluated at the scale $\mu = H$ and computed in the ${\overline{{\rm MS}}}$ scheme to next-to-next-to-leading order precision \cite{Degrassi:2012ry}. Lower panels show the same for the fermion channel for different values of $y_{\psi}$ with $\xi=0.0262$ and $\lambda=10^{-5}$, including the case with $y_{\psi}$ equal to the SM top quark Yukawa coupling $y_{\rm t}(\mu = H)$  \cite{Degrassi:2012ry}. The values of $\xi$ and $\lambda$ are chosen such that they maximise the gravitational wave amplitude for the largest $g$ and $y_{\psi}$ values shown in the figure. 

The ultraviolet cutoff in the figures marks the region above which our computation of ${\cal P}_{\zeta_{\chi}}(k)$ is no longer applicable. Modes above the cutoff correspond to length scales below the coarse-graining scale chosen in our $\delta N$ computation. They could be studied by choosing a smaller coarse-graining scale and properly accounting for the time evolution of $\zeta$ during their horizon crossing when the reheating has not yet fully completed and isocurvature perturbations are still present. The resulting physical spectrum should be smoothly decaying above our sharp ultraviolet cutoff but since these scales are well above the observationally testable gravitational wave frequencies, we do not investigate this issue in further detail here.

\begin{figure}[t!]
	\begin{subfigure}[b]{0.32\textwidth}
		\includegraphics[width=\linewidth]{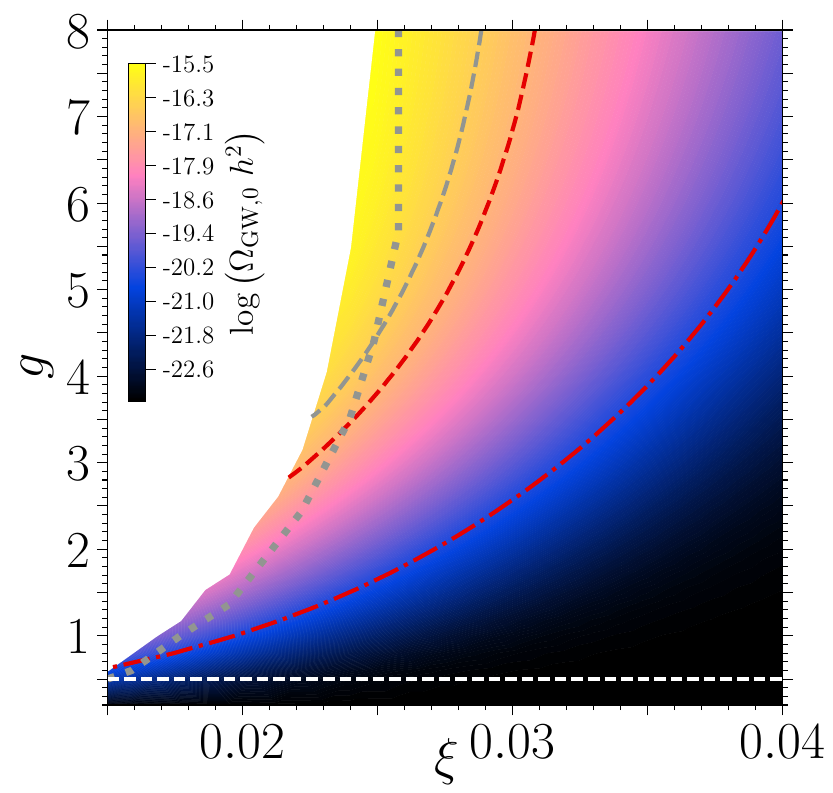}
		\label{fig:vectorGW-5}
	\end{subfigure}
	\begin{subfigure}[b]{0.32\textwidth}
		\includegraphics[width=\linewidth]{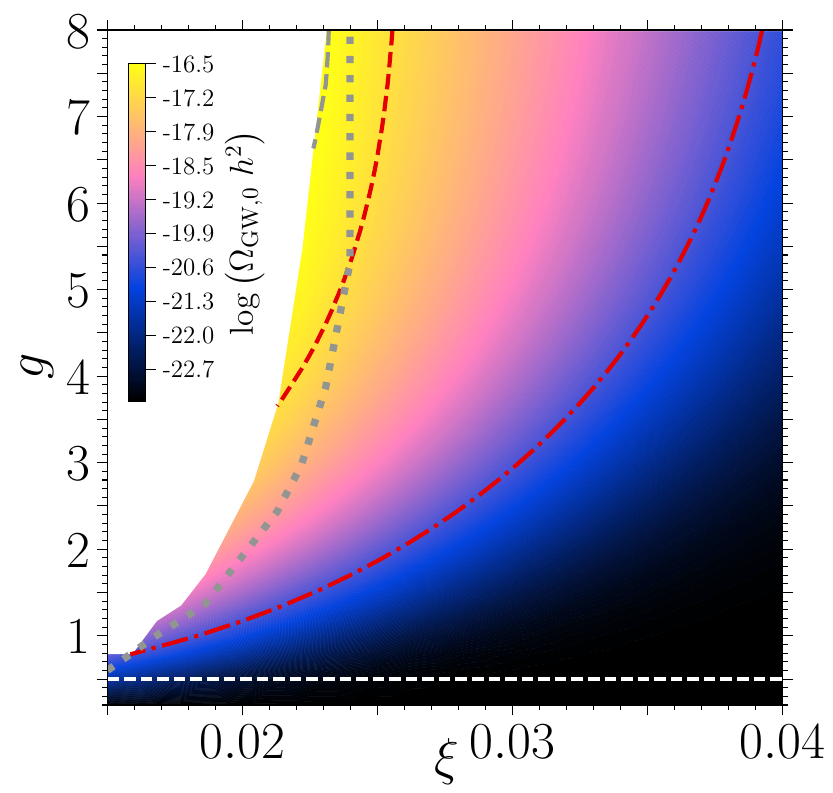}
		\label{fig:vectorGW-3}
	\end{subfigure}
	\begin{subfigure}[b]{0.32\textwidth}
		\includegraphics[width=\linewidth]{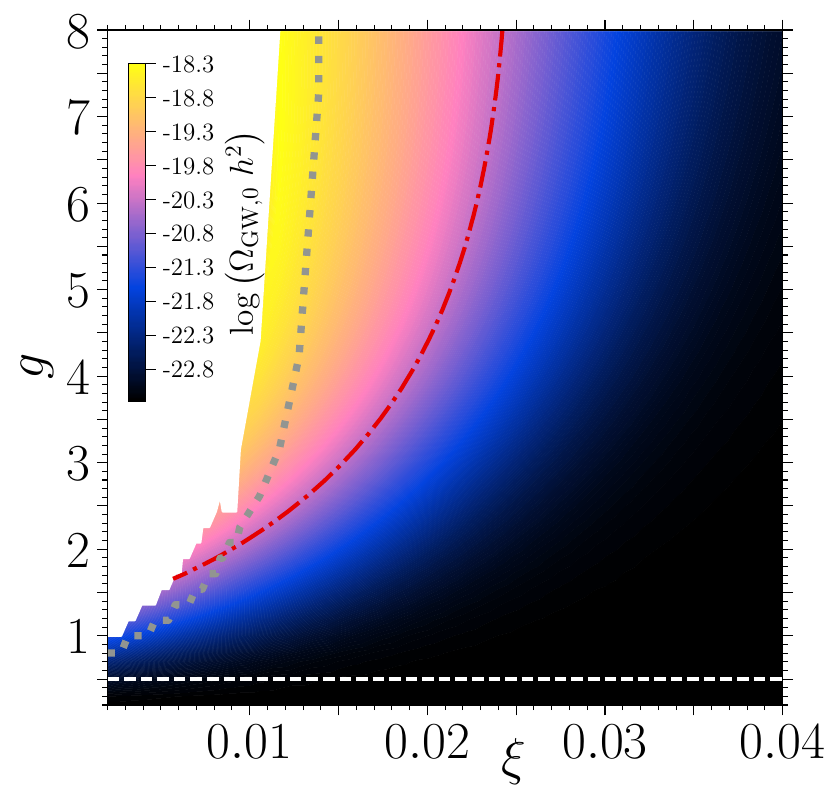}
		\label{fig:vectorGW-1}
	\end{subfigure}
	\centering
	\begin{subfigure}[b]{0.32\textwidth}
		\includegraphics[width=\linewidth]{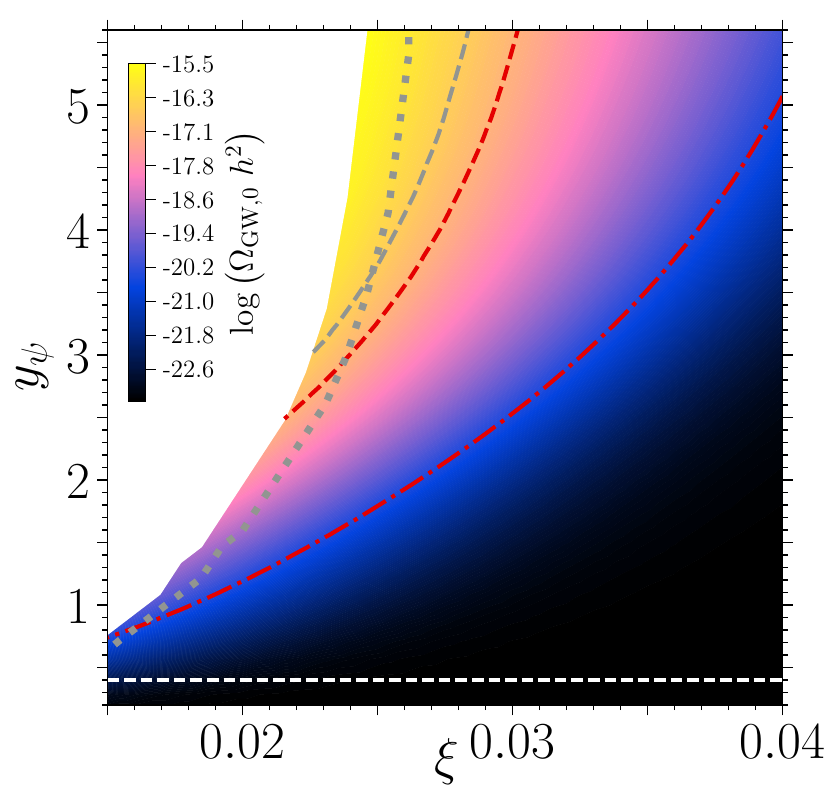}
		\label{fig:fermionGW-5}
	\end{subfigure}
	\begin{subfigure}[b]{0.32\textwidth}
		\includegraphics[width=\linewidth]{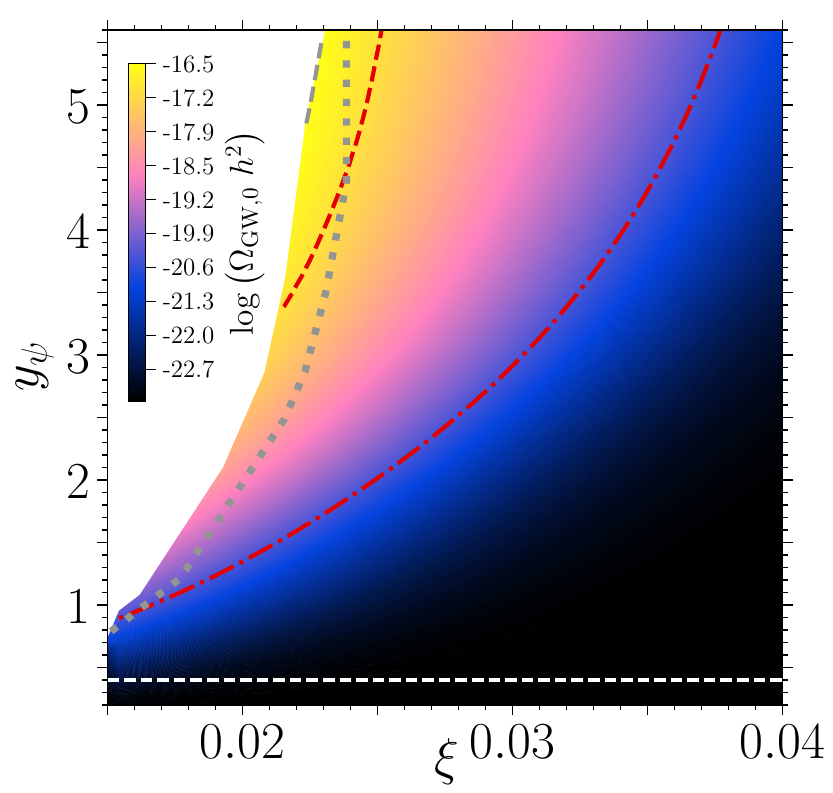}
		\label{fig:fermionGW-3}
	\end{subfigure}
	\begin{subfigure}[b]{0.32\textwidth}
		\includegraphics[width=\linewidth]{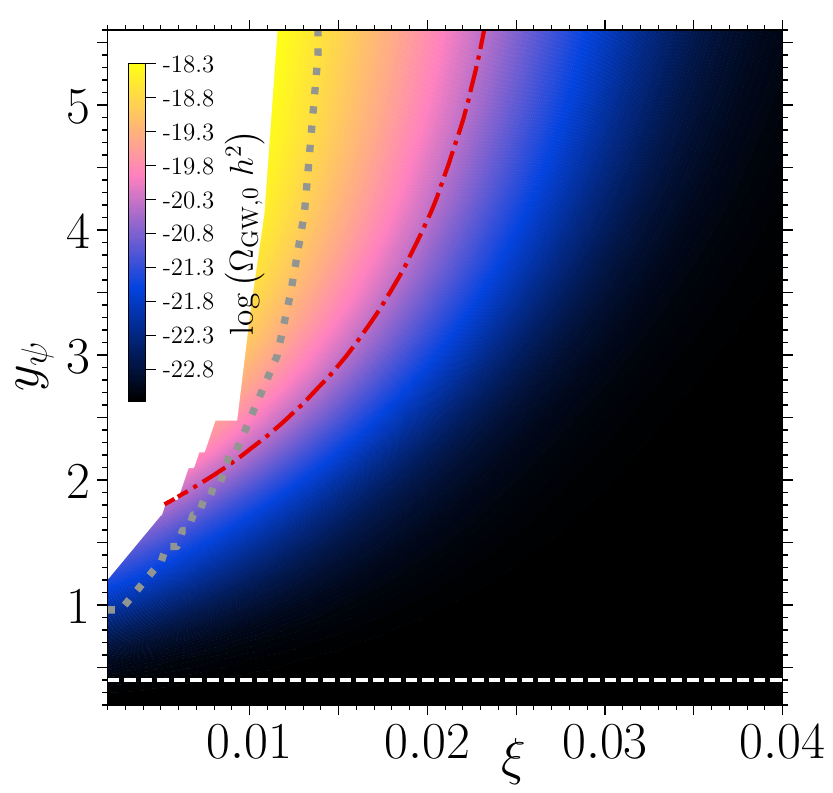}
		\label{fig:fermionGW-1}
	\end{subfigure}
	
	\caption{The top row shows the GW spectral density today at $f=0.15~{\rm Hz}$ as function of $\xi$ and $g$ for $\lambda=10^{-5},\,10^{-3},\,10^{-1}$ (from left to right). The bottom row shows the same as function of $\xi$ and $y_\psi$.  The (conservative) non-Gaussianity bound $P_{\zeta_\chi}(k_*)  < 10^{-12}$ is violated to the left of the grey dashed lines and in the white region $P_{\zeta_\chi}(k_*) > 5 \times 10^{-12}$. The white dotted lines show results for the SM coupling values as in Figure~\ref{fig:curves}. The power-law integrated sensitivity curves with SNR $=1$ are shown for BBO (red dashed), DECIGO (grey dashed) and Ultimate DECIGO (red dashed-dotted). 
}
	\label{fig:GW_scan}
\end{figure}

Figure~\ref{fig:GW_scan} shows the gravitational wave density fraction at the frequency $f=0.15~{\rm Hz}$ as function of $\xi$ and $g$ for three different values of $\lambda$. The relevant experimental sensitivity curves are plotted with the same line styles as in Figure~\ref{fig:curves}.  The grey dotted curve denotes the line ${\cal P}_{\zeta_\chi}(k_*) = 1 \times 10^{-12}$ and to the right of this curve the (conservative) non-Gaussianity bound (\ref{eq:pchibound}) is satisfied. The boundary of the white region corresponds to ${\cal P}_{\zeta_\chi}(k_*) = 5 \times 10^{-12}$ for which $f_{\rm NL}^{\rm max}$ obtained in Appendix~\ref{app:B} is still marginally below the Planck 1$\sigma$ bound for all couplings $\xi < 0.04$. 

As can be seen in Figures \ref{fig:curves} and \ref{fig:GW_scan}, apart from the region $g, y_{\psi} \gg 1$, the gravitational waves produced in our setup appear to be unobservably small in the entire $\lambda,\xi$ range we have studied. 
For $g, y_{\psi} \gg 1$, the gravitational wave signal could be potentially detectable by surveys like BBO or DECIGO but, as discussed above, in this region it is highly questionable if the spectator sector (\ref{eq:action}) can be considered as a perturbative high-energy limit of any nearly ultraviolet complete SM extension. We reiterate that for  this reason our results in the region $g, y_{\psi} \gg 1$ should primarily be understood as results of a phenomenological setup where the inflaton decay rate is just parameterised by equations (\ref{eq:gamma1}) and (\ref{eq:gamma2}).

The main factor that suppresses the gravitational wave signal is the non-Gaussianity constraint (\ref{eq:pchibound}).  It is however not immediately obvious how this large-scale bound translates to gravitational waves since they probe the spectrum on much smaller scales and ${\cal P}_{\zeta_{\chi}}$  can be strongly blue-tilted.  The spectral tilt in equation (\ref{eq:Pzetachi}) is determined by the couplings $\xi$ and $\lambda$. Increasing their values makes the spectrum more blue-tilted but simultaneously also tends to decrease the amplitude ${\cal P}_{\zeta_{\chi}}(k_*)$.  The decrease in the amplitude can be compensated by increasing the couplings $g$ and $y_{\psi}$ which do not affect the spectral tilt. Increasing $g$ and $y_{\psi}$ leads to stronger modulation of the reheating dynamics and therefore larger ${\cal P}_{\zeta_{\chi}}(k_*)$ but the dependence is quite convoluted due to details of the modulated reheating dynamics \cite{Karam:2020skk}. As a result, it is a non-trivial parameteric question how large ${\cal P}_{\zeta_{\chi}}(k)$, and $\Omega_{\rm GW,0}$, can be on scales relevant for the gravitational wave surveys while still being compatible with the bispectrum bound (\ref{eq:pchibound}) on large scales, and the outcome is shown in Figure \ref{fig:GW_scan}. 

Finally, the gravitational wave amplitude also depends on the couplings $\Lambda_{(1,2)}$ which determine the overall amplitude of the inflaton decay rates (\ref{eq:gamma1}) and (\ref{eq:gamma2}). The values chosen in Figures \ref{fig:curves} and \ref{fig:GW_scan} nearly maximise the decay rates, and therefore the gravitational wave signal, while still being compatible with modelling the decay rates by equations (\ref{eq:gamma1}) and (\ref{eq:gamma2}) in the entire parameter range shown in the figures. Decreasing the values of $\Lambda_{(1,2)}$ further by a factor of 2 or so makes the decay rates dynamically relevant already slightly before the onset of inflaton oscillations where the use of equations (\ref{eq:gamma1}) and (\ref{eq:gamma2}) would be questionable.

\section{Conclusions}

We have investigated scalar-induced stochastic gravitational waves sourced by a spectator field through the modulated reheating mechanism. The spectator scalar in our setup has a Higgs-like potential with a self-coupling  $\lambda$ and a non-minimal coupling $\xi$ to spacetime curvature, and the inflaton decay proceeds perturbatively through shift-symmetric dimension-five operators to vectors or fermions \cite{Lu:2019tjj}. The vector and fermion masses are determined by the spectator field, similar to the mass terms generated by the Higgs field, and spatial fluctuations in the spectator value are converted to adiabatic curvature perturbations over the reheating epoch. 

In our analysis, we have assumed that the spectator has relaxed to the de Sitter vacuum during inflation and identified ensemble expectation values in this state as observable volume averages. Under these premises, the curvature perturbation sourced by the spectator is strongly non-Gaussian and has to be subdominant on scales probed by the CMB anisotropies. However, the spectrum of the spectator sourced curvature perturbation ${\cal P}_{\zeta_{\chi}}(k)$ is blue-tilted and its amplitude can be much larger on small scales $k \ll {\rm Mpc}^{-1}$. For example, if  $\xi \gtrsim 0.02$ or $\lambda \gtrsim 0.3$, the spectrum ${\cal P}_{\zeta_{\chi}}(k)$ grows at least by a factor $10^5$ from the CMB pivot scale $k_{*} = 0.05 {\rm Mpc}^{-1}$ to $k= 10^{14} {\rm Mpc}^{-1}$, which is approximatively in the middle of the frequency range probed by future  gravitational wave surveys like BBO and DECIGO. It is therefore a priori possible that the spectator perturbations could source  detectable gravitational wave signals.   

In this work, we have applied the stochastic formalism of inflation and the spectral expansion method to compute the non-linear power spectrum ${\cal P}_{\zeta_{\chi}}(k)$ and standard second order perturbation theory methods to determine the gravitational wave signal induced by it. We have also computed the bispectrum using a truncated expression for $\zeta_{\chi}$ and found that the Planck constraints on non-Gaussianity imply a bound ${\cal P}_{\zeta_{\chi}}(k_*) < 10^{-12}$. Imposing this bound, we have scanned over a broad range of spectator parameter space and computed the gravitational wave density fraction. 

The gravitational wave amplitude generated in the setup depends both on the couplings $\xi$ and $\lambda$ in the effective potential of the spectator and on the couplings $g$ and $y_{\psi}$ that appear in the vector and fermion channel mass terms. 
Our results indicate that observable gravitational waves cannot be generated in the regime $g, y_{\psi} \lesssim 1$.    
For large coupling values, $g, y_{\psi} \gtrsim 3$, the setup can produce  $\Omega_{\rm GW, 0}h^2 \gtrsim 10^{-17}$ at $f \sim 0.1 {\rm Hz} $ which would be marginally detectable with BBO or DECIGO. 
For such large couplings, it is however difficult to consider the spectator setup as a nearly ultraviolet complete extension of the SM perturbatively extrapolated up to the inflationary energy scale and realised below eventual Landau poles. 
Therefore, one of our main outcomes is that within the modulated reheating framework studied in this work it appears that stochastic gravitational waves cannot be used as a probe of the Higgs sector, or other spectator scalars in SM extensions.

A possible caveat in this conclusion is the de Sitter vacuum assumed in our work. First, even if the spectator is in the vacuum, volume averages over patches of the size of the observable universe can differ from the ensemble averages. A non-vanishing volume average $\langle \chi\rangle_{\rm V}$ would generate a small Gaussian term in $\zeta_{\chi}$ and relax the non-Gaussianity bound ${\cal P}_{\zeta_{\chi}}(k_*) < 10^{-12}$, allowing for somewhat larger gravitational wave amplitudes. We have not performed a systematic quantitative study of such finite volume effects but we note that the potentially interesting regime for gravitational waves in our setup is characterised by $\xi = {\cal O}(0.01)$ which implies that the spectator effective mass during inflation is not negligible, $m_{\rm eff}^2/H^2 \sim 0.1 $. The mass decreases the spectator correlation length and therefore also suppresses impacts of random long-wavelength fluctuations.  Second, if the spectator is significantly displaced from vacuum during inflation, the ensemble expectation value $\langle\chi\rangle$ is non-vanishing and  $\zeta_{\chi}$ can be Gaussian to leading approximation. This is the so-called mean field limit commonly assumed in spectator setups which are constructed to generate a sizeable fraction of the Gaussian and nearly scale-invariant large-scale perturbations, and where the model predictions depend in the input value of the classical spectator field. In the mean field limit, the Planck constraints on non-Gaussianity are expected to place much less stringent constraints on ${\cal P}_{\zeta_{\chi}}(k_*)$ and the gravitational wave signal could therefore be significantly stronger than in our current setup.  

\acknowledgments

We thank David Weir for helpful comments on  gravitational wave sensitivity curves. The financial support from the Research Council of Finland (grant\# 342777) is gratefully acknowledged.
 
\appendix

\section{Two-point functions of $\zeta_{\phi}$ and $\zeta_{\chi}$}
\label{app:A}

Using equations (\ref{eq:zetafull}), (\ref{eq:zetaphi}) and (\ref{eq:zetachi}), and setting $\langle \delta\phi \chi^n\rangle = 0$, the two point function of $\zeta$ on superhorizon scales becomes  
\beq
\langle\zeta({\bf x})\zeta({\bf x'})\rangle = \langle\zeta_{\phi}({\bf x})\zeta_{\phi}({\bf x'})\rangle+\langle\zeta_{\chi}({\bf x})\zeta_{\chi}({\bf x'})\rangle~,
\eeq
where
\baq
\label{eq:appazp2}
\langle\zeta_{\phi}({\bf x})\zeta_{\phi}({\bf x'})\rangle  &=&\langle \partial_{\phi}N(\bar{\phi}_{\rm i},\chi_{\rm i}({\bf x}))\partial_{\phi}N(\bar{\phi}_{\rm i},\chi_{\rm i}({\bf x}'))\rangle \langle \delta\phi_{\rm i}({\bf x})\delta\phi_{\rm i}({\bf x}')\rangle~, 
\\ 
\label{eq:appazc2}
\langle\zeta_{\chi}({\bf x})\zeta_{\chi}({\bf x'})\rangle  &=& \langle N(\bar{\phi}_{\rm i},\chi_{\rm i}({\bf x}))N(\bar{\phi}_{\rm i},\chi_{\rm i}({\bf x}'))\rangle-\langle N(\bar{\phi}_{\rm i},\chi_{\rm i}({\bf x}))\rangle^2~.
\eaq
Here $\bar{\phi}_{\rm i} \equiv \bar{\phi}(t_{\rm i})$, $\chi_{\rm i}({\bf x})\equiv \chi(t_{\rm i},{\bf x})$ and $\delta\phi_{\rm i}({\bf x})\equiv \delta\phi(t_{\rm i},{\bf x})$, and the initial time $t_{\rm i}$ can be chosen freely. In the main text we determined the initial time via $\epsilon_{\rm H}(t_{\rm i}) =0.1$ but here we do not yet make this choice. In the following $t_{\rm i}$ is an unspecified initial time unless explicitly stated otherwise. 

The spectrum of the inflaton perturbations $\langle \delta\phi({\bf k}) \delta\phi({\bf k'}) \rangle = (2\pi)^3 \delta({\bf k}+{\bf k}')(2\pi^2/k^3){\cal P}_{\delta\phi}(k) $ in the spatially flat gauge and on superhorizon scales is given by the standard first order perturbation theory result for single field slow roll inflation \cite{Mukhanov:1990me}
\beq
\label{eq:Pdeltaphi}
{\cal P}_{\delta\phi}(t_{\rm i},k)=\left(\frac{H_{\rm i}}{2\pi}\right)^2\left(1+2 {\rm ln}\left(\frac{a_{\rm i}H_{\rm i}}{k}\right)(3\epsilon-\eta)\right)~.
\eeq
In this expression the coefficients of the $k$-dependent and $k$-independent parts are separately expanded to leading order precision in slow roll.  The slow roll parameters are defined in the usual way, $\epsilon = M_{\rm P}^2/2(V'/V)^2$ and $\eta = M_{\rm P}^2V''/V$ and evaluated at $t_{\rm i}$.

We use the stochastic formalism and the spectral expansion approach to compute the two-point correlators involving the spectator field in equations (\ref{eq:appazp2}) and (\ref{eq:appazc2}). We approximate the spectator distribution using the de Sitter vacuum solution and set $H=H_{\rm i}$.
In the de Sitter vacuum, the joint equal time two-point distribution for 
$\chi(t_{\rm i},{\bf x})$ can be written as \cite{Starobinsky:1994bd,Markkanen:2019kpv,Markkanen:2020bfc} 
\beq
\label{eq:rho2}
\rho_{2}(\chi,{\bf x},t_{\rm i};\chi',{\bf x}',t_{\rm i}) =\psi_{0}(\chi)\psi_{0}(\chi')\sum_{n=0}^{\infty}\psi_n(\chi)\psi_n(\chi')\left(a_{\rm i}H_{\rm i} \Delta x \right)^{-2\Lambda_n/H_{\rm i}}~,
\eeq
where  $\Delta x \equiv |{\bf x}-{\bf x}'|$.  
The eigenfunctions $\psi_n(\chi)$ and eigenvalues $\Lambda_n$ are determined by 
\beq
\psi_n''(\chi) + \left(U''(\chi)-U'(\chi)\right) \psi_n(\chi)= -\frac{4\pi^2 \Lambda_n}{H_{\rm i}^3}\psi_{n}(\chi)~,
\eeq
where $U(\chi)$ is given by equation (\ref{eq:Uchi}) and boundary conditions are set as $\psi_n(\chi) \rightarrow 0 $ for $\chi\rightarrow \pm\infty$. The eigenfunctions are orthonormal with 
\beq
\label{eq:AppAorthonormal}
\int_{-\infty}^{\infty}{\rm d}\chi \psi_n(\chi)\psi_m(\chi) = \delta_{nm}~.
\eeq

Using equations (\ref{eq:appazp2}) and (\ref{eq:rho2}), the two-point function of the inflaton sourced part $\zeta_{\phi}$ becomes  
\baq
\label{eq:AppA_zphitwopoint}
\langle\zeta_{\phi}({\bf x})\zeta_{\phi}({\bf x'})\rangle
&=&\langle 0| N'|0\rangle^2 \langle\delta\phi_{\rm i}({\bf x})\delta\phi_{\rm i}({\bf x}')\rangle\left(1 +\sum_{n=1}^{\infty}\langle 0| N' |2 n\rangle^2 \left(a_{\rm i}H_{\rm i} \Delta x\right)^{\frac{-2\Lambda_n}{H_{\rm i}}} \right)~,~~~~~~~~~
\eaq
where $N'\equiv \partial_{\phi}N(\bar{\phi}_{\rm i},\chi_{\rm i}({\bf x}))$ and we have denoted
\beq
\langle 0| f(\chi) |n\rangle \equiv \int_{-\infty}^{\infty}{\rm d}\chi \psi_0(\chi)f(\chi)\psi_n(\chi)~.
\eeq
We have also used that $N(\bar{\phi}_{\rm i},\chi_{\rm i})$ is an even function of $\chi_{\rm i}$. Since we consider slow-roll inflation and in our setup the spectator field does not affect the dynamics of the universe before the end of inflation, $t_{\rm end}$, the number of e-folds can be written as 
\beq
N(\bar{\phi}_{\rm i},\chi_{\rm i}({\bf x})) =  \int_{\bar{\phi}_{\rm i}}^{\bar{\phi}_{\rm end}}{\rm d} {\bar{\phi}} \frac{H(\bar{\phi})}{\dot{\bar{\phi}}}+\int_{t_{\rm end}}^{t_{\rm f}} {\rm d} t H(\bar{\phi}(t),\chi_{\rm i}({\bf x}))~.
\eeq 
The first term is the usual slow-roll expression and the second term depends on $\chi_{\rm i}({\bf x})$ that affects the reheating history. Only the first term depends on $\bar{\phi}_{\rm i}$ and contributes to equation (\ref{eq:AppA_zphitwopoint}). Since it does not depend on $\chi_{\rm i}$ and the eigenfunctions obey the orhonormality relation (\ref{eq:AppAorthonormal}), equation (\ref{eq:AppA_zphitwopoint}) reduces to the standard slow-roll result  
\baq
\label{eq:AppA_zphitwopoint0}
\langle\zeta_{\phi}({\bf x})\zeta_{\phi}({\bf x'})\rangle
&=&\left(\frac{H(\bar{\phi}_{\rm i})}{\dot{\bar{\phi}}_{\rm i}}\right)^2 \langle \delta\phi({\bf x})\delta\phi({\bf x}')\rangle  ~.
\eaq
Taking the Fourier transform, using equation (\ref{eq:Pdeltaphi}), and setting $t_{\rm i}$ equal to the horizon crossing time $t_k$ of a mode $k$, defined via $k = a(t_k)H(t_k)$,  we obtain the usual expression (\ref{eq:Pzetaphi}) for the spectrum.  

Using equations (\ref{eq:appazc2}) and (\ref{eq:rho2}) we can write the two-point function of $\zeta_{\chi}$ as 
\baq
\label{eq:AppA_zetachi2ptsum}
\nonumber
\langle\zeta_{\chi}({\bf x})\zeta_{\chi}({\bf x'})\rangle
&=&\int_{-\infty}^{\infty} {\rm d}\chi_{\rm i} \rho_{2}(\chi_{\rm i},{\bf x},t_{\rm i};\chi_{\rm i}',{\bf x}',t_{\rm i}) N(\bar{\phi}_{\rm i},\chi_{\rm i})N(\bar{\phi}_{\rm i},\chi_{\rm i}')
\\\nonumber
&&-\int_{-\infty}^{\infty} {\rm d}\chi_{\rm i} \rho_{2}(\chi_{\rm i},{\bf x},t_{\rm i};\chi_{\rm i},{\bf x},t_{\rm i}) N(\bar{\phi}_{\rm i},\chi_{\rm i})^2
\\
&=&\sum_{n=1}^{\infty}\langle 0| N(\bar{\phi}_{\rm i},\chi_{\rm i}) |n \rangle^2 (a_{\rm i} H_{\rm i} \Delta x)^{-\frac{2\Lambda_n}{H_{\rm i}}}~. 
\eaq
Taking the Fourier transform, we obtain the spectrum given by equation (\ref{eq:Pzetachi}) in the text. 

\section{Bispectrum of $\zeta_{\chi}$}
\label{app:B}

The curvature perturbation in (\ref{eq:zetachi}) is an even function of $\chi$ and it can be expanded as   
\beq
\label{eq:appzetaexp}
\zeta_{\chi}({\bf x}) = \frac{1}{2}N''(\chi_{\rm i}({\bf x})^2-\langle\chi_{\rm i}^2\rangle)+\sum_{n=2}^{\infty}\frac{1}{(2n)!}N^{(2n)}(\chi_{\rm i}({\bf x})^{2n}-\langle\chi_{\rm i}^{2n}\rangle)~,
\eeq
where $N^{(n)}\equiv \partial^{n} N/\partial\chi^n$ are evaluated at $\chi=0$.

Here we focus on the limit where the non-minimal coupling $\xi R \chi^2$ dominates over the self-coupling $\lambda \chi^4$ during inflation, generating a quadratic effective potential with the mass $m^2 = 12 \xi H_{\rm i}^2$. In this limit, $\chi_{\rm i}({\bf x})$ is a Gaussian field and the infrared limit of the two-point function is given by the standard Bunch-Davies result
\beq
\label{eq:BD}
\langle \chi_{\rm i}({\bf x})\chi_{\rm i}({\bf x}') \rangle  = \left(\frac{H_{\rm i}}{2\pi}\right)^2\frac{\Gamma(\nu)\Gamma(\frac{3}{2}-\nu)}{\sqrt{\pi}}
(a_{\rm i} H_{\rm i}|{\bf x}-{\bf x}'|)^{-(3-2\nu)}~,~~\nu=\sqrt{\frac{9}{4}-12\xi}~.
\eeq
We further assume that $(N^{(2n)}/N'')\langle \chi_{\rm i}^{2n-2}\rangle \ll 1$ and truncate the series (\ref{eq:appzetaexp}) at the first term 
\beq
\label{eq:appzeta}
\zeta_{\chi}({\bf x}) = \frac{1}{2}N''(\chi_{\rm i}({\bf x})^2-\langle\chi_{\rm i}^2\rangle)~.
\eeq

The bispectrum of the three-point function of the total curvature perturbation $\zeta$ is defined in the momentum space as 
\beq
\langle\zeta({\bf k}_1)\zeta({\bf k}_2)\zeta({\bf k}_3)\rangle = (2\pi)^3 \delta({\bf k}_1+{\bf k}_2+{\bf k}_3) B({\bf k}_1,{\bf k}_2,{\bf k}_3)~.
\eeq
Neglecting slow-roll suppressed contributions from $\zeta_{\phi}$, the bispectrum is entirely due to the spectator component $\zeta_{\chi}$. A straightforward computation gives   
\beq
\label{eq:bispectrum1}
B({\bf k}_1,{\bf k}_2,{\bf k}_3) = N''{}^3 C_{\chi}^3 (a_{\rm i} H_{\rm i})^{-(9-6\nu)}I({\bf k}_1,{\bf k}_2,{\bf k}_3)~,
\eeq
where $C_{\chi}$ is a constant given by
\beq
\label{eq:appb_c}
C_{\chi}=\left(\frac{H_{\rm i}}{2\pi}\right)^2 2^{2\nu}\pi \Gamma(\nu)^2~,
\eeq
and $I$ is a convolution integral given by
\beq
\label{eq:I}
I({\bf k}_1,{\bf k}_2,{\bf k}_3) = \int \frac{{\rm d}{\bf q}}{(2\pi)^3}|{\bf q}|^{-2\nu}|{\bf q}-{\bf k}_2|^{-2\nu}|{\bf q}+{\bf k}_1|^{-2\nu}~.
\eeq
The integral has no ultraviolet or infrared divergences for $1/2 < \nu < 3/2$. This covers the entire $\xi$ range relevant in our work and we focus on this interval in what follows. 

Note that the bispectrum sourced by a Gaussian squared type spectator component has been investigated already in  \cite{Boubekeur:2005fj} but approximating the integral (\ref{eq:I}) with methods that yield a final result which depends on a cutoff scale imposed by hand. Here we evaluate the full integral (\ref{eq:I}) without resorting to approximative methods and our final result involves no arbitrary cutoff scales.     

To proceed, we introduce the Feynman parameters and rewrite equation (\ref{eq:I}) as 
\beq
\label{eq:Ifeyn}
I = \frac{\Gamma(3\nu)}{\Gamma(\nu)^3}\int \frac{{\rm d}{\bf q}}{(2\pi)^3}\int_0^1 {\rm d} u_1 \int_0^1 {\rm d} u_2 \int_0^1 {\rm d} u_3 \frac{\delta(1-u_1-u_2-u_3) (u_1 u_2 u_3)^{\nu-1}}{(u_1 |{\bf q}|^2+u_2 |{\bf q}-{\bf k}_2|^2 +u_3 |{\bf q}+{\bf k}_1|^2)^{3 \nu}}~.
\eeq
By shifting ${\bf q} \rightarrow {\bf q}+u_2 {\bf k}_2-u_3 {\bf k}_1$, the ${\bf q}$ integral can be separated from the rest as
\beq
\label{eq:Ishifted}
I  = \frac{\Gamma(3\nu)}{\Gamma(\nu)^3}\int_0^1 {\rm d} u_1 \int_0^1 {\rm d} u_2 \int_0^1 {\rm d} u_3 \delta(1-u_1-u_2-u_3) (u_1 u_2 u_3)^{\nu-1}\int \frac{{\rm d}{\bf q}}{(2\pi)^3} (|{\bf q}|^2+\mu^2)^{-3\nu}~,
\eeq
where $\mu^2 = u_2|{\bf k}_2|^2+u_3|{\bf k}_1|^2-|u_2 {\bf k}_2-u_3 {\bf k}_1|^2 \geqslant 0$. Performing the ${\bf q}$ integral and one of the Feynman parameter integrals  we get an intermediate result
\beq
\label{eq:Iintermediate}
I  = \frac{\Gamma(3\nu-\frac{3}{2})}{8\pi^{\frac{3}{2}}\Gamma(\nu)^3}\int_0^1 {\rm d} u_2 \int_0^{1-u_2} {\rm d} u_3 ((1-u_2-u_3) u_2 u_3)^{\nu-1}(\mu^2)^{\frac{3}{2}-3\nu}~.
\eeq
After a change of variables $u_3 \rightarrow u_3/(1-u_2)$, the integral over $u_2$ can also be performed analytically and we get
\beq
\label{eq:Ifinal}
I  = \frac{|{\bf k}_1|^{3-6\nu}\Gamma(3\nu-\frac{3}{2})\Gamma(\frac{3}{2}-\nu)}{8\pi^2 \Gamma(\nu)^2 (2+\nu(6\nu-7))}\int_0^1 {\rm d} u \left(f_1(u,\kappa,\theta) - 8\left(\nu-\frac{1}{2}\right)f_2(u,\kappa,\theta)\right)~,
\eeq
where $\kappa = |{\bf k}_2|/|{\bf k}_1|$, ${\rm cos}\theta = {\bf k}_1\cdot {\bf k}_2/(|{\bf k}_1||{\bf k}_2|)$, and 
\baq
\nonumber
f_1&=& (u(1-u))^{\frac{1}{2}-2\nu}\left(1+\frac{u(1-u)}{\kappa^2+u^2+2 u {\rm cos}\theta}\right)\left[ {}_2F_1\left(\nu, -\frac{3}{2}+3\nu, -\frac{1}{2},-\frac{\kappa^2+u^2+2u\kappa {\rm cos}\theta}{u(1-u)}\right) \right.
\\
&&
\left.- {}_2F_1\left(\nu, -\frac{3}{2}+3\nu, \frac{1}{2},-\frac{\kappa^2+u^2+2u\kappa {\rm cos}\theta}{u(1-u)}\right)\right]~,
\\
f_2&=&(u(1-u))^{\frac{1}{2}-2\nu} {}_2F_1\left(\nu, -\frac{3}{2}+3\nu, \frac{1}{2},-\frac{\kappa^2+u^2+2u\kappa {\rm cos}\theta}{u(1-u)}\right)~.
\eaq
Here  ${}_2F_1$ is the hypergeometric function. 

The remaining integral in equation (\ref{eq:Ifinal}) can be carried out numerically. To this end we note theta both $f_1$ and $f_2$ contain integrable singularities at the integral limits. To perform the numerical integrals, we first expand $f_1$ and $f_2$ around $u=0$ and $u=1$, subtract the singular terms from  $f_1$ and $f_2$ and integrate numerically over the remaining regular parts. Then we analytically integrate over the singular terms and add the results together. 
\begin{figure}[t!]
\centering
\includegraphics[width=0.5 \textwidth]{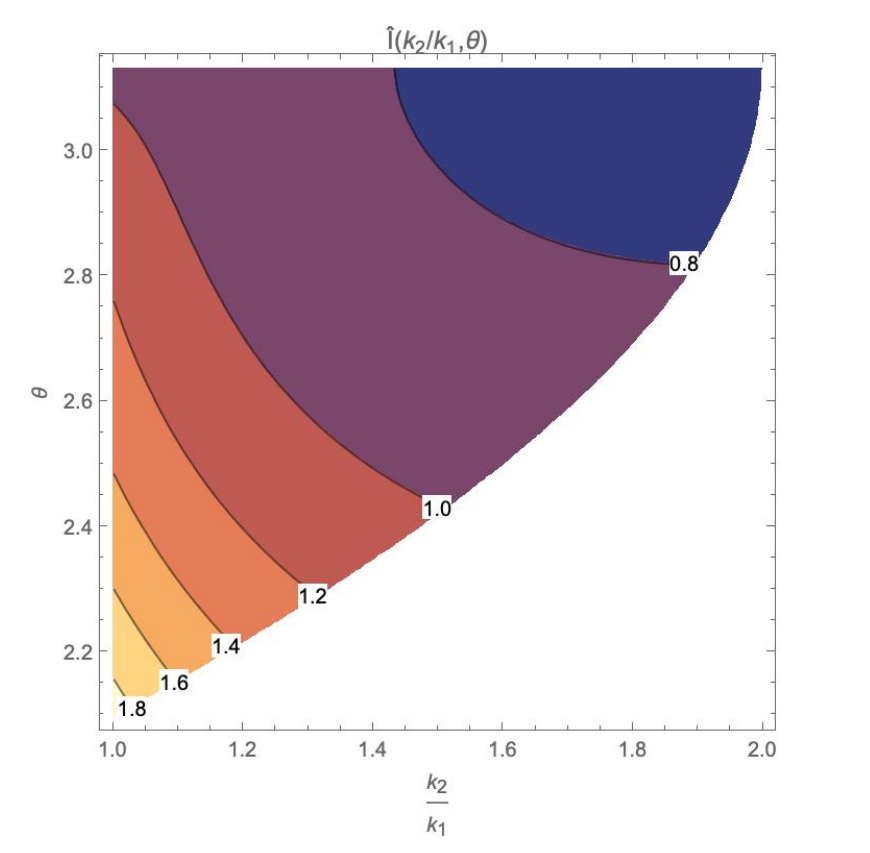}
\caption{The function $\hat{I}(\kappa,\theta)$ in the bispectrum. Here $\kappa = k_2/k_1$, ${\rm cos}\theta = {\bf k}_1\cdot {\bf k}_2/(k_1 k_2)$ and the momenta are labelled such that $k_{3} \leqslant k_{1}\leqslant k_{2}$. We have set the non-minimal coupling $\xi = 0.01$ which corresponds to $\nu = 1.46$. Configurations in the white region do not satisfy the momentum conservation condition ${\bf k}_1+{\bf k}_2+{\bf k}_3=0$.}       
\label{fig:ihat}
\end{figure}

We choose the label the momenta such that $k_{3} \leqslant k_{1}\leqslant k_{2}$. In the squeezed limit, $k_3\ll k_1\sim k_2$, we find that $I({\bf k}_1,{\bf k}_2,{\bf k}_3)$ given by equation (\ref{eq:Ifinal}) diverges approximatively as $k_3^{(3-6\nu)/2}$ for the $\xi$ range investigated in our spectator setup. Defining a dimensionless quantity 
\baq
\label{eq:Ihat}
\hat{I}(\kappa,\theta) &=&  \left(\kappa^2+1+2\kappa{\rm cos}\theta \right)^{-(3+6\nu)/4}\frac{\Gamma(3\nu-\frac{3}{2})\Gamma(\frac{3}{2}-\nu)}{8\pi^2 \Gamma(\nu)^2 (2+\nu(6\nu-7))}\times
\\\nonumber
&&
\int_0^1 {\rm d} u \left(f_1(u,\kappa,\theta) - 8\left(\nu-\frac{1}{2}\right)f_2(u,\kappa,\theta)\right)~,
\eaq
and using that $\kappa^2+1+2\kappa{\rm cos}\theta = k^2_3/k^2_1$, the bispectrum (\ref{eq:bispectrum1}) can be written as  
\beq
\label{eq:bIhat}
B({\bf k}_1,{\bf k}_2,{\bf k}_3) = N''{}^3 C_{\chi}^3 (a_{\rm i} H_{\rm i})^{-(9-6\nu)}(k_1 k_3)^{(3-6\nu)/2}\hat{I}(\kappa,\theta)~.
\eeq
The function $\hat{I}(\kappa,\theta)$ is depicted in Figure \ref{fig:ihat} which shows only a weak dependence on the wavenumbers.  Similar weak dependence is observed over the entire range of $\xi$ values investigated in this work, which corresponds to $3/2-\nu \lesssim 0.2$. We therefore conclude that the bispectrum in our setup scales nearly  as $k_{3}^{-3}$ in the squeezed limit $k_3\ll k_1\sim k_2$ and therefore approximatively corresponds to the local type of bispectrum \cite{Komatsu:2001rj}.
\begin{figure}[t!]
\centering
\includegraphics[width=0.45 \textwidth]{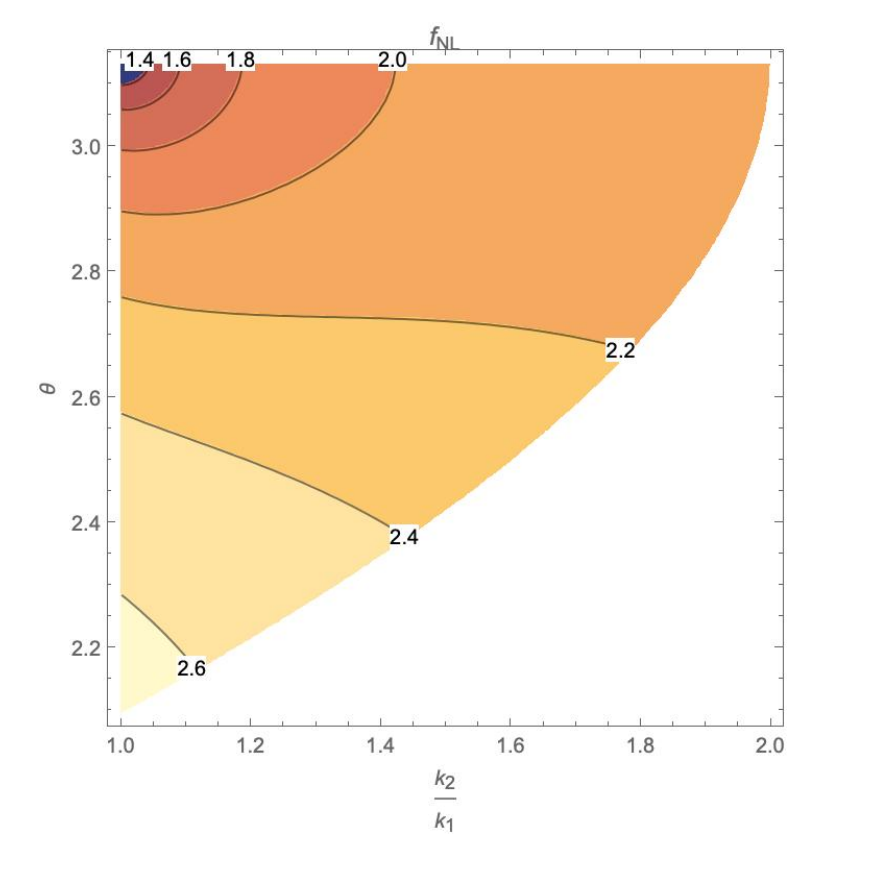}
\caption{The scale dependent non-linearity parameter $f_{\rm NL}$ shown as function of the ratio $\kappa = k_2/k_1$ and the angle $\theta$ between ${\bf k}_1$ and ${\bf k}_2$. The momenta are labelled such that $k_{3} \leqslant k_{1}\leqslant k_{2}$ and we have set $k_2 = 0.2 \, {\rm Mpc}^{-1}$. The other parameters are set as $\xi = 0.01$ ($\nu = 1.46$) and ${\cal P}_{\zeta_{\chi}}(k_*)/{\cal P}_{\zeta}(k_*) = 10^{-11}$ at $k_* =0.05\, {\rm Mpc}^{-1}$. Configurations in the white region do not satisfy the momentum conservation condition ${\bf k}_1+{\bf k}_2+{\bf k}_3=0$.}       
\label{fig:fnl}
\end{figure}

In order to compare the bispectrum (\ref{eq:bIhat}) with the observational constraints, we recast it in terms of the power spectrum. The power spectrum of the total curvature perturbation $\zeta = \zeta_{\phi} + \zeta_{\chi}$ on CMB scales can be written in the usual way in terms of the amplitude and the spectral index 
\beq
\label{eq:appbpzeta}
\langle \zeta({\bf k})\zeta({\bf k'})\rangle = (2\pi)^3 \delta({\bf k}+{\bf k}') P_{\zeta}(k) = (2\pi)^3 \delta({\bf k}+{\bf k}') \frac{2\pi^2}{k^3}{\cal P}_{\zeta}(k_*)\left(\frac{k}{k_*}\right)^{n_{\rm s}-1}~,
\eeq
where $k_* = 0.05 {\rm Mpc}^{-1}$ is the Planck pivot scale \cite{Planck:2018vyg}. Using equations (\ref{eq:BD}) and (\ref{eq:appzeta}) we find the dimensionless power spectrum of $\zeta_{\chi}$ given by  
\beq
\label{eq:appbpzetachi}
{\cal P}_{\zeta_{\chi}} (k) = N''{}^2 C_{\chi}^2 
\frac{\Gamma(\frac{3}{2}-\nu)^2\Gamma(-\frac{3}{2}+2\nu)}{2^5\pi^{7/2}\Gamma(\nu)^2\Gamma(3-2\nu)}
\left(\frac{k}{a_{\rm i}H_{\rm i}}\right)^{2(3-2\nu)}~,
\eeq
for $3/4 < \nu <3/2$. Using equations (\ref{eq:appbpzeta}) and (\ref{eq:appbpzetachi}), we can recast (\ref{eq:bIhat}) as
\baq
\label{eq:bfinal}
B({\bf k}_1,{\bf k}_2,{\bf k}_3) &=& P_{\zeta}(k_1) P_{\zeta}(k_3) \left(32\sqrt{2}\pi^{5/4} \left(\frac{\Gamma(\nu)^2\Gamma(3-2\nu)}{\Gamma(\frac{3}{2}-\nu)^2\Gamma(-\frac{3}{2}+2\nu)}\right)^{\frac{3}{2}}{\cal P}_{\zeta}(k_*)^{-\frac{1}{2}}\times\right.~~~~
\\\nonumber 
&&
\left.\left(\frac{{\cal P}_{\zeta_{\chi}}(k_*)}{{\cal P}_{\zeta}(k_*)}\right)^{\frac{3}{2}}\left(\frac{k_1k_3}{k_*^2}\right)^{\frac{3}{2}(3-2\nu)-(n_{\rm s}-1)}\hat{I}(\kappa,\theta)\right)~,
\eaq
where we recall that the momenta are labeled such that $k_{3} \leqslant k_{1}\leqslant k_{2}$. 
\begin{figure}[t!]
\centering
\includegraphics[width=0.5 \textwidth]{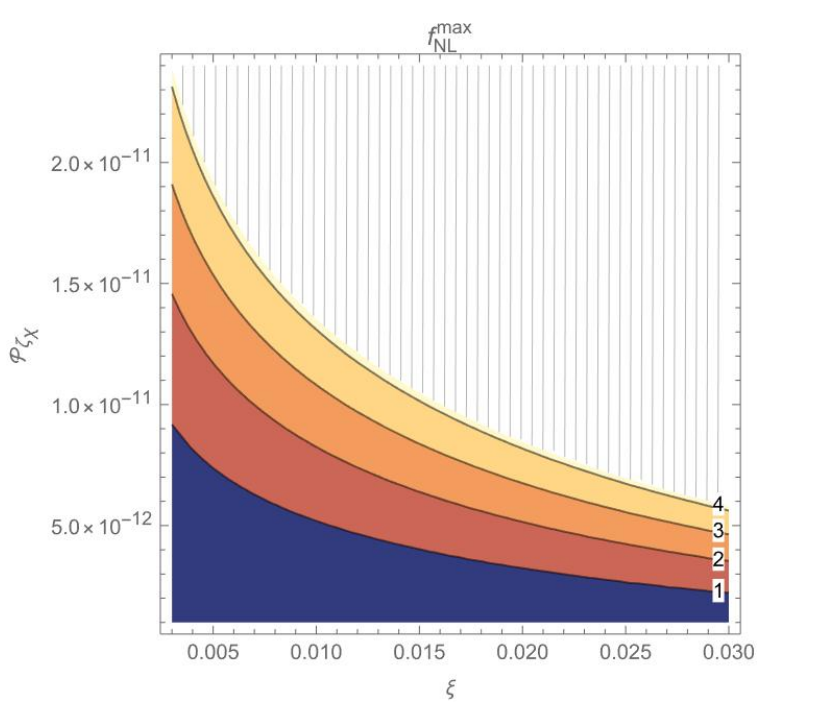}
\caption{The maximum value of the shape dependent non-linearity parameter $f^{\rm max}_{\rm NL}$ shown as function of $\xi$ and ${\cal P}_{\zeta_{\chi}}(k_*)$ where $k_* =0.05\, {\rm Mpc}^{-1}$. The hatched regions corresponds to $f^{\rm max}_{\rm NL} >4.2$ which is outside the $1\sigma$ region of  Planck constraint on local non-Gaussianity $f^{\rm local}_{\rm NL} = -0.9 \pm 5.1$.}       
\label{fig:fnlmax}
\end{figure}

As discussed above, the bispectrum (\ref{eq:bfinal}) is approximatively of the local type. Following \cite{Byrnes:2009pe,Byrnes:2010ft} we define the scale-dependent local non-linearity parameter as  
\beq
\label{eq:fnldef}
f_{\rm NL}({\bf k}_1,{\bf k}_2,{\bf k}_3) = \frac{5}{6}\frac{B({\bf k}_1,{\bf k}_2,{\bf k}_3)}{P_{\zeta}(k_1) P_{\zeta}(k_3)+P_{\zeta}(k_2) P_{\zeta}(k_3)+P_{\zeta}(k_1) P_{\zeta}(k_2)}~.
\eeq
Using equation (\ref{eq:bfinal}) we obtain 
\baq
\label{eq:fnlfinal}
f_{\rm NL}(\kappa,\theta, k_2) &=& \frac{80}{3}\sqrt{2}\pi^{5/4} \left(\frac{\Gamma(\nu)^2\Gamma(3-2\nu)}{\Gamma(\frac{3}{2}-\nu)^2\Gamma(-\frac{3}{2}+2\nu)}\right)^{\frac{3}{2}}{\cal P}_{\zeta}(k_*)^{-\frac{1}{2}}\left(\frac{{\cal P}_{\zeta_{\chi}}(k_*)}{{\cal P}_{\zeta}(k_*)}\right)^{\frac{3}{2}}\times
\\\nonumber
&&
\hat{I}(\kappa,\theta)\left(\kappa^2+1+2 \kappa {\rm cos}\theta\right)^{\frac{3}{2}\left(\frac{3}{2}-\nu\right)+1-n_{\rm s}}\left(\frac{k_2}{k_*}\right)^{3(3-2\nu)+2(1-n_{\rm s})}\kappa^{2+n_{\rm s}-3(3-2\nu)}\times
\\\nonumber
&&
\left(\left(\kappa^2+1+2 \kappa {\rm cos}\theta\right)^{2-\frac{n_{\rm s}}{2}}+\kappa^{4-n_{\rm s}}+1\right)^{-1}~.
\eaq

For our convention of labelling the wavenumbers, $k_{3} \leqslant k_{1}\leqslant k_{2}$, the momentum configuration of the bispectrum is determined by the largest wavenumber $k_2$, and the shape parameters $\kappa$ and $\theta$. The dependence of $f_{\rm NL}(\kappa,\theta, k_2)$ on its arguments is illustrated in figure \ref{fig:fnl}. As can be seen in the figure, $f_{\rm NL}$ has a relatively mild dependence on $\kappa$ and $\theta$, and it is  maximised for $\kappa = 1$ and $\theta = 2\pi/3$ which corresponds to the equilateral limit. According to equation (\ref{eq:fnlfinal}), $f_{\rm NL}$ scales as  $k_2^{3(3-2\nu)+2(1-n_{\rm s})}$ and is therefore a growing function of $k_2$. 

The Planck analysis on scale-dependent local non-Gaussianity in \cite{Planck:2019kim} does not include a template directlty applicable for equation (\ref{eq:fnlfinal}).  We will therefore just compare our results against the Planck constraint on local non-Gaussianity $f^{\rm local}_{\rm NL} = -0.9 \pm 5.1$ \cite{Planck:2019kim}. In order to get conservative constraints, we use the maximal value of (\ref{eq:fnlfinal}) in the comparison. For the range of wavenumbers $ k \lesssim 0.2 {\rm Mpc}^{-1}$ probed by Planck, equation (\ref{eq:fnlfinal}) is  maximised for 
\beq
f_{\rm NL}^{\rm max}\equiv f_{\rm NL}(k_2 = 0.2 {\rm Mpc}^{-1}, \kappa = 1, \theta = 2\pi/3)~. 
\eeq  
The value of $f_{\rm NL}^{\rm max}$ is depicted in Figure \ref{fig:fnlmax} as function of $\xi$ and ${\cal P}_{\zeta_{\chi}}(k_*)$.

\section{On the numerical integrals}
\label{app:numericsGW}

\begin{figure}[h]
           \centering
            \includegraphics[width=0.4\linewidth]{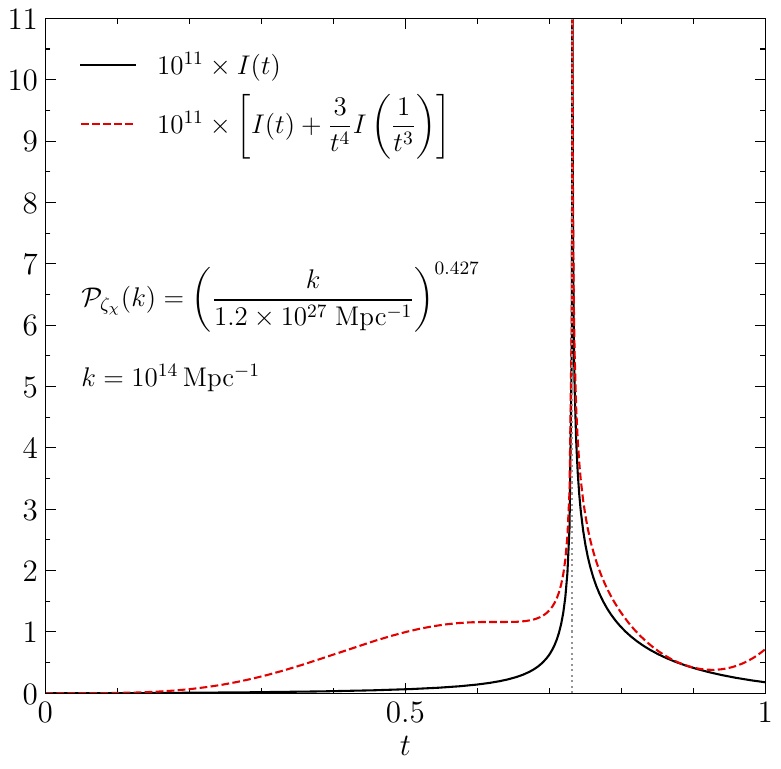}
        \caption{
The integrand $I(t)$ of equation ~(\ref{eq:Omega_GW_P_zeta_expression}) (black line) and its rescaled form (red dashed line) given by equation (\ref{eq:identity_I}).  The spectrum, ${\cal P}_{\zeta_\chi}(k)$, corresponds to the $g=8$ curve in figure~\ref{fig:curves}. 
        }
    \label{fig:I_vs_rescaled_I}
    \end{figure}

The expression for the gravitational wave density fraction   \eqs{eq:Omega_GW_P_zeta_expression} has an integrable singularity at $t=\sqrt{3} -1$. To improve the convergence of the numerical integration, we rescale the integration limits over $t$ by applying the identity 
    \begin{equation}
    	\int_{0}^{\infty} dt I(t) = \int_{0}^{1} dt \lrsb{I(t)+\dfrac{n}{t^{n+1}}I\lrb{\dfrac{1}{t^n}}} \;,
    	\label{eq:identity_I}
    \end{equation}
for the integrand $I(t)$ of equation ~(\ref{eq:Omega_GW_P_zeta_expression}). As can be seen in Figure~\ref{fig:I_vs_rescaled_I}, this smears the integrand around  $t=\sqrt{3} -1$ and therefore enables our integration method to adapt its grid faster. In the numerical results we present, we have used the {\tt python} module {\tt vegas} that implements the {\tt VEGAS+} algorithm~\cite{Lepage:1977sw,Lepage:2020tgj}. 

\bibliography{modreh_GW}
\bibliographystyle{JHEP-edit}

\end{document}